\documentclass[iop,apj,numberedappendix,appendixfloats,twocolappendix]{emulateapj}
\usepackage{apjfonts}
\usepackage{amsmath}
\usepackage{iondefs}
\usepackage{symbols}

\slugcomment{Accepted to ApJ January 2015}
\shorttitle{\sc CGM Analysis of Simulations}
\shortauthors{\sc Churchill {\etal}}

\begin{document}

\title{Direct Insights into Observational Absorption Line Analysis Methods \\
  of the Circumgalactic Medium Using Cosmological Simulations}

\author{\sc
Christopher W. Churchill\altaffilmark{1},
Jacob R. Vander Vliet\altaffilmark{1},
Sebastian Trujillo-Gomez\altaffilmark{1}, \\
Glenn G. Kacprzak\altaffilmark{2}
and 
Anatoly Klypin\altaffilmark{1}
}

\altaffiltext{1}{New Mexico State University, MSC 4500, Las Cruces, NM 88003,
USA}
\altaffiltext{2}{Centre for Astrophysics and Supercomputing, Swinburne
University of Technology, PO Box 218, Victoria 3122, Australia}


\begin{abstract}
We study the circumgalactic medium (CGM) of a $z=0.54$ simulated dwarf
galaxy using hydroART simulations. We present our analysis methods,
which emulate observations, including objective absorption line
detection, apparent optical depth (AOD) measurements, Voigt profile
(VP) decomposition, and ionization modeling.  By comparing the
inferred CGM gas properties from the absorption lines directly to the
gas selected by low ionization {\HI} and {\MgII}, and by higher
ionization {\CIV} and {\OVI} absorption, we examine how well
observational analysis methods recover the ``true'' properties of CGM
gas.  In this dwarf galaxy, low ionization gas arises in
sub-kiloparsec ``cloud'' structures, but high ionization gas arises in
multiple extended structures spread over 100 kpc; due to complex
velocity fields, highly separated structures give rise to absorption
at similar velocities.  We show that AOD and VP analysis fails to
accurately characterize the spatial, kinematic, and thermal conditions
of high ionization gas.  We find that {\HI} absorption selected gas
and {\OVI} absorption gas arise in totally distinct physical gas
structures, calling into question current observational techniques
employed to infer metallicities and the total mass of ``warm-hot'' CGM
gas.  We present a method to determine whether {\CIV} and {\OVI}
absorbing gas is photo or collisionally ionized and whether the
assumption of ionization equilibrium is sound.  As we discuss, these
and additional findings have strong implications for how accurately
currently employed observational absorption line methods recover the
true gas properties, and ultimately, our ability to understand the CGM
and its role in galaxy evolution.
\end{abstract}


\section{Introduction}
\label{sec:intro}

A challenge placed before $\Lambda$CDM hydrodynamic cosmological
simulations is to form realistic galaxies while matching such
quantities as the stellar-mass to halo-mass relation and average star
formation history as a function of halo mass and redshift
\citep[e.g.,][]{behroozi10, behroozi13, moster13, munshi13,
  st13-dwarfs, Ceverino14, agertz14}.  In the simulations, the
interplay between stellar feedback processes, originating in the
interstellar medium (ISM), and filamentary and galaxy accretion,
originating in the intergalactic medium (IGM), give rise to extended
metal-enriched gaseous structures surrounding galaxies, i.e., the
circumgalactic medium (CGM).  The exact role of the CGM in governing
the observed properties of galaxies is not yet well established;
however, it is quite possible that the CGM is a highly regulating
component of galaxies and, if better understood, could provide
powerful insights into global galaxy relations
\citep[cf.,][]{oppenheimer10, mathes14}

Thus, it is an additional challenge for successful simulations to also
statistically match the observed distributions of gas density,
temperature, kinematics, and chemical and ionization conditions of the
CGM.  In general, there are two main approaches to simulating galaxies
and the CGM.  The first uses smoothed particles hydrodynamics (SPH),
which discretizes gas mass into particles, and the second uses
adaptive mesh refinement (AMR), which spatially discretizes the gas
using grid cells.

SPH simulations generally trade off high spatial and mass resolution
in favor of modeling the hydrodynamics in thousands of galaxies in a
simulation box.  A great strength of SPH simulations is that the
statistical characteristics of the CGM can be studied over a wide
range of halo mass and cosmological environment.  The draw back is
that the detailed physics of stellar formation and feedback are
generally in the form of scaling relations, such as constant velocity
winds, or winds with launch velocities proportional to the stellar
velocity dispersion, which scales with the gravitational potential
\citep[e.g.][]{oppenheimer08, freeke12}.  These relations do not
directly address the underlying physics of feedback.

A strength of AMR simulations is that the star formation and feedback
models directly target the underlying physical processes of star
formation and feedback.  Though the processes are still unresolved,
the physics is highly detailed and the simulations can be employed to
study star formation at the scale of molecular cloud physics and
stellar feedback at the scale of radiation pressure physics and
photo-heating physics \citep{st13-dwarfs, Ceverino14, agertz14}.  With
such detail, the draw back is that only a few galaxies can be
simulated at a time.  However, with AMR, greater insight into the
complex interplay between star formation and feedback that regulates
the CGM, and therefore galaxy formation and evolution, can be gleaned.

Observationally, the primary method for studying the gas properties of
the CGM is the technique of quasar absorption lines.  The most
commonly studied CGM absorption lines are {\HI}~$\lambda 1215$
({\Lya}) and {\HI}~$\lambda 1025$ ({\Lyb}) \citep[e.g.,][]{lanzetta95,
  stocke13, tumlinson13, mathes14}, the lithium isosequence zero-volt
resonant doublets {\CIVdblt} and {\OVIdblt} \citep[e.g.,][]{simcoe06,
  stocke06, fox07-ovi, fox07-civ, tumlinson11, stocke13, mathes14},
and the sodium isosequence zero-volt resonant doublet {\MgIIdblt}
\citep[e.g.,][and references therein]{nielsen-cat1, nielsen-cat2}.
Virtually all of the physical conditions we have learned about the CGM
are derived from absorption line analysis.  As such, studying the CGM
properties of simulated galaxies and quantitatively comparing these
properties to those derived from observations is clearly best
accomplished using absorption line measurement and analysis methods
\citep[also see][]{ford13a, ford13b, hummels13}.

Absorption line analysis of simulations will not only directly
increase our knowledge of the CGM and its role in galaxy evolution,
but will provide a profound insight into how effective and accurate
commonly applied observational analysis methods recover the ``true''
gas properties (densities, temperatures, chemical and ionization
conditions, and kinematics).  To accomplish these objectives, we must
quantitatively compare the inferred gas properties derived from
observational techniques applied to synthetic absorption lines in
simulations directly to the physical properties of the simulation gas
from which the absorption arises.

For example, ionization models \citep[e.g., Cloudy,][]{ferland98,
  ferland13} are used to determine the gas density, $n\subH$, and
metallicity, as well as the ionization parameter, $U=
n_{\gamma}/n\subH$, where $n_\gamma$ is the number density of hydrogen
ionizing photons.  Many assumptions are invoked in applying these
models, such as single-phase absorbing clouds
\citep[e.g.,][]{cwc-q1317, lehner13, stocke13, fox14, werk14}, or
simple two-phase absorbing clouds \citep[e.g.,][]{cc99, charlton03,
  ding03, tripp11, ggk-q1317}.  How appropriate are these assumptions?
Quantifying the relationships between absorption profiles, the
inferred properties of the absorbing gas from the absorption line
data, and the true'' properties of the gas giving rise to the
absorption in the simulations can provide insight to this question.

A second example is the inferred kinematics, which is either assumed
to be provided by the profile of the column density with velocity via
the apparent optical depth (AOD) column density method
\citep[e.g.,][]{savage91}, or reflected in Voigt profile (VP)
decomposition of the absorption profile \citep[e.g.,][]{boksy79,
  cvc03, simcoe06, mathes14}.  However, inferring the properties of
the gas using either the AOD profile method and VP decomposition
implicitly relies on the assumption that gas at a given velocity
arises from a single unique spatial location along the line of sight.
Furthermore, VP decomposition models the data as isothermal clouds,
each having a different peculiar velocity.  Using simulations, the
relative spatial locations of the absorbing gas along the LOS can be
examined to quantify the degree to which absorption with similar (or
aligned) kinematics arises in the same spatial gas structures.  This
latter information has important implications for the assumptions
applied to the ionization modeling and kinematic analysis of observed
data.

To be effective, the application of absorption line analysis to
simulations should emulate observational work as closely as possible
so that the selection methods, spectral resolution, sensitivity
limitations, and analysis techniques of various absorption-line
surveys can be accurately duplicated.  In general, surveys target a
limited number of absorption line transitions, each which probe a
relatively narrow range of gas phase, i.e., they arise in gas with
favorable density, temperature, ionization conditions, and
metallicities.  This will also be true for absorption lines in
synthetic spectra obtained by passing sightlines through the CGM in
hydrodynamic simulations.  For example, along a single sightline, some
gas structures may contribute to {\MgII} and/or {\CIV} absorption,
whereas other unique gas structures may contribute to {\OVI}
absorption, but not to {\MgII} or {\CIV} absorption.

But, importantly, all our knowledge of the CGM from absorption line
analysis is filtered through the instruments that modify the data.
Which instrument and telescope facility is used to observe a given
absorption line depends upon the rest wavelength of the transition and
the redshift of the absorbing structure, and this strongly governs the
design, sensitivity, and data quality for observational surveys.
Since which absorption lines get observed depend upon which facilities
capture the wavelength range of the redshifted transition, this
governs which gas phases can be probed and to what level of
sensitivity they can be probed.

Furthermore, the range of physical gas properties that contribute to
detectable absorption will differ as a function of the detection
thresholds of the spectra, which depends on the signal-to-noise ratio,
resolution, and pixelization of the data.  Shallower detection
thresholds result in probing gas with higher column densities, which
presumably means that higher density, higher metallicity gas, or
favorable ionization conditions are preferentially being studied.
This holds true for both real-world observations and synthetic spectra
from simulations, provided that the synthetic spectra are carefully
designed to emulate real-world spectra.

\begin{figure*}
\epsscale{0.9}
\plotone{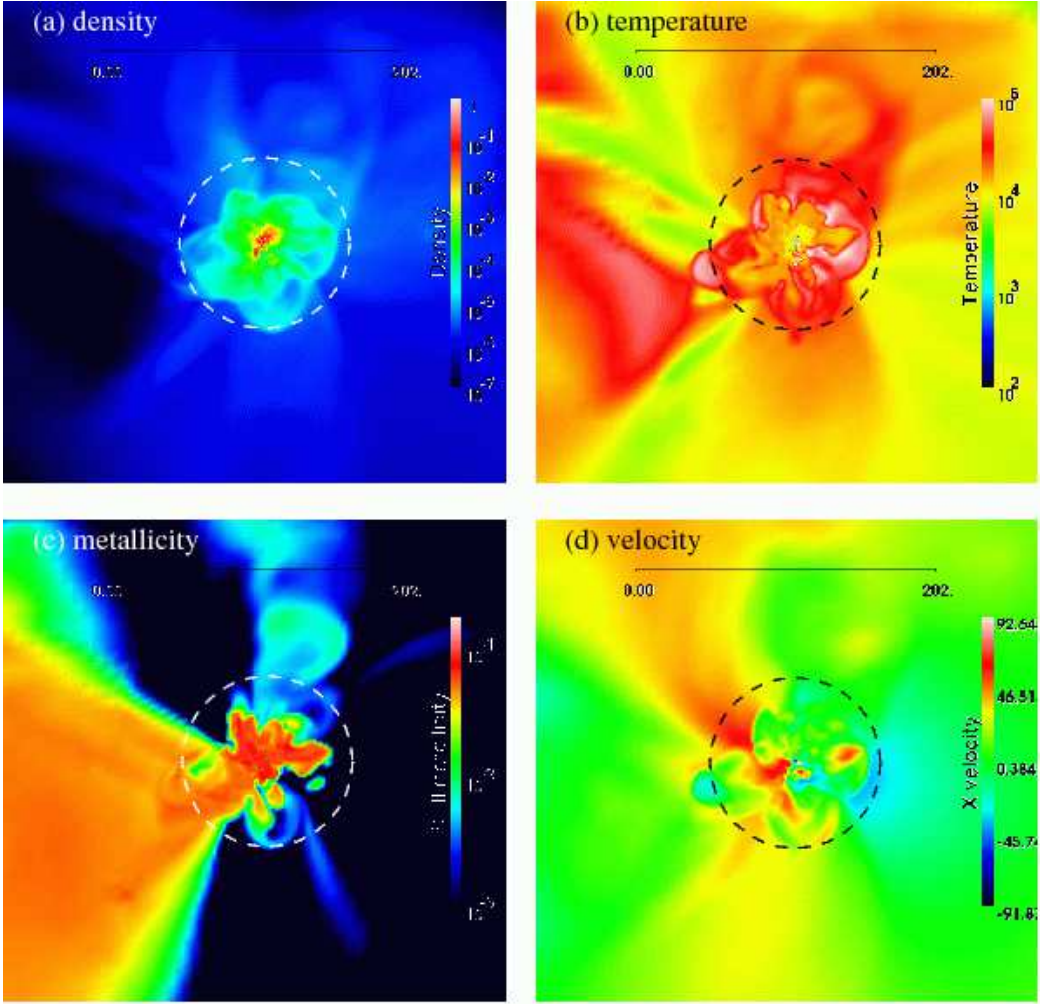}
\caption{Slices centered on the center of mass of a $z=0.54$ simulated
  galaxy for the dwALL\_8 feedback model of \citet{st13-dwarfs},
  showing the gas (a) number density, (b) temperature, (c) Type II
  supernovae metal mass fraction yield, and (d) $x$-component of the
  velocity (where the $x$ direction is positive to the right).  The
  horizontal bar across each panel provides the physical scale in
  proper kiloparsecs and the dashed circles show the virial radius.
  The dwALL\_8 star formation and feedback model includes
  deterministic star formation in molecular cloud environments,
  supernova heating, stellar winds, radiation pressure, and
  photo-heating.  At $z=0.54$, the galaxy has virial mass $M_{\rm vir}
  = 2.7 \times 10^{10}$ M$_{\odot}$, virial radius $R_{\rm vir} = 57$
  kpc, and stellar mass $M_{\ast} = 1.2 \times 10^7$ M$_{\odot}$.}
\label{fig:simshots}
\end{figure*}

With the aforementioned considerations in mind, in this paper we
describe the methods we developed to generate ``realistic'' synthetic
spectra through simulated galaxies in AMR cosmological simulations and
to analyze the spectra, all in a manner that emulates real-world
observations.  In Section~\ref{sec:sims} we describe the simulated
galaxy we employ for this study.  In Section~\ref{sec:ionmodel}, we
briefly review the ionization model \citep{cwc-ioncode} used to obtain
the gas ionization conditions.  We detail our methods for absorption
line analysis in Section~\ref{sec:simabslines} and discuss selected
preliminary findings in Section~\ref{sec:discuss}.  In
Section~\ref{sec:conclude}, we summarize our work.

\section{The Simulations}
\label{sec:sims}


We employ the $N$-body plus gasdynamics AMR code hydroART
\citep{kravtsov97, kravtsovPhDT, kravtsov99, kravtsov04}. The
simulations were run using the ``zoom-in'' technique \citep{klypin01},
which allows us to resolve the formation of a single galaxy
consistently in its full cosmological context. The high-resolution
region around the galaxy is typically $\sim 1$--2 Mpc across and the
hydrodynamics is resolved with $7 \times 10^6$ grid cells, with a
minimum cell size of roughly $30~h^{-1}$ pc at $z=0$ (the proper size
of the cells decrease with increasing redshift).

Physical processes implemented in the code include star formation,
stellar feedback, Type II and Ia metal enrichment, thermal and
radiation pressure, and metallicity-dependent cooling and heating.
Gas is self-shielded, advects metals, is heated by a homogeneous
ultraviolet background, and can cool to 300 K due to metal and
molecular line cooling.  Gas flows, shock fronts, and metal
disbursement follow self-consistently from this physics.  

For star formation, we use observations of molecular clouds
\citep{KrumholzTan07} to guide our model.  The star particles form in
the dense, cold molecular phase ($n_{\hbox{\tiny H}} \sim 100$ {\cc},
$T \simeq 100$ K).  The star formation rate is proportional to the gas
density divided by the free fall time of the molecular cloud.  We use
an observationally motivated low (2\% percent) efficiency per free
fall time for converting gas into stars.  Compared to previous methods
\citep{ceverino09, Ceverino10}, this treatment results in a greater
number of individual star particles, but with smaller masses, between
$\sim 100$ and $\sim 10^3$ M$_{\odot}$.  Runaway young, hot stars are
included according to \citet{ceverino09} by providing one third of the
newly-formed star particles with a random velocity kick.

For the feedback model, mechanical energy from stellar winds and SN
type II is assumed to thermalize and is injected into the gas as
thermal energy around young stars following the rates predicted by
Starburst99 \citep{Leitherer99} for the Chabrier IMF
\citep{chabrier03}.  We also incorporated photoionization heating,
radiation pressure, and shocked stellar winds from massive stars
\citep[see][for details]{ceverino09, Ceverino10, st13-dwarfs,
  Ceverino14}. These feedback processes disrupt the cold molecular gas
and regulate the formation of stars \citep{st13-dwarfs, Ceverino14}.

Matching observations of {\HII} regions \citep{Lopez13}, we treat
photoheating by adding a non-thermal pressure ($P/k \simeq 10^{6}$ K
cm$^{-3}$) to the gas surrounding young stars. This pressure decreases
rapidly in order to reproduce the declining density within a growing
{\HII} region.  For radiation pressure, which we treat similarly to
\citet{Murray10}, \citep{Hopkins11}, and \citet{Agertz12}, we include
momentum from the radiation field from young massive, which is coupled
to the gas and dust through scattering and absorption.  Absorption of
UV photons scales as $1 - e^{-\tau_{\hbox{\tiny UV}}} \simeq 1$.
Scattering due to trapped IR photon scales as $\tau_{\hbox{\tiny IR}}
\simeq 1$.  These values are adopted from the suggested values of
\citep{KrumholzThompson12} based upon $\tau_{\hbox{\tiny IR}}$ scaling
with column density \citep{Davis14}.

This star formation and feedback model was shown to reproduce many
properties of low-mass galaxies at $z \sim 0$ without fine tuning
\citep[see][]{st13-dwarfs}; the stellar to halo mass ratio, cold
gas fraction, baryon content, star formation history, rotation curves
and morphologies of the simulated galaxies agree remarkably well with
observations.

The heating and cooling balance of the gas is determined using heating
and cooling functions obtained from Cloudy \citep{ferland98,
  ferland13} and incorporates equilibrium photoionization and
collisional ionization.  These include metal and molecular line
cooling, and a uniform \citet{haardt01} ionizing background with
self-shielding of high column density gas, and stellar radiation for
gas in the vicinity of stars.  For additional details see
\citet{Ceverino10}.

At each grid cell, the hydroART code follows the evolution of the
density, temperature, velocity, and metal mass fraction. The metals
produced in type II and Ia supernovae are followed separately and are
self-consistently advected with the gas flow. To compute the relative
abundances of the ions in the gas, we employ an equilibrium ionization
model as a post-processing step. We briefly describe the ionization
model in Section~\ref{sec:ionmodel}; full details are given in
\citet{cwc-ioncode}.

For the analysis in this paper, we adopt the simulation of the
low-mass (dwarf) galaxy designated dwALL\_8 presented in
\citet{st13-dwarfs}.  For this simulation, the host dark matter halo
evolved into an isolated dwarf galaxy with a virial mass $M_{\rm vir}
= 2.8 \times 10^{10}$ M$_{\odot}$, a virial radius $R_{\rm vir} = 80$
kpc, a stellar mass $M_{\ast} = 2.1 \times 10^7$ M$_{\odot}$, and a
maximum circular velocity $v_c = 60$ {\kms} at $z=0$.  As described
above, the dwALL\_8 model incorporates feedback in which the optical
depth of the gas and dust to IR photons is small and the pressure of
the gas due to photo-heating in {\HII} regions is $P/k = 8 \times
10^6$ K cm$^{-3}$.  These assumptions are within the range favored by
observations of star forming regions.  The dark matter particle
resolution is $9.4 \times 10^4$ M$_{\odot}$.


We study the galaxy when it is at redshift $z=0.54$.  At this
redshift, the galaxy has virial mass $M_{\rm vir} = 2.7 \times
10^{10}$ M$_{\odot}$, virial radius $R_{\rm vir} = 57$ kpc, and
stellar mass $M_{\ast} = 2.1 \times 10^7$ M$_{\odot}$.  In the galaxy
interstellar medium, the minimum gas cell proper size is $27~h^{-1}$
pc.  In the circumgalactic medium, the proper cell sizes range from
$27~h^{-1}$ pc to $432~h^{-1}$ pc.  Within two virial radii, there are
$\sim 7 \times 10^6$ gas cells. 

In Figure~\ref{fig:simshots}, we present thin slices through the
gas distribution centered on the galaxy at $z=0.54$, showing (a) gas
density, (b) temperature, (c) metallicity, and (d) the $x$-component
of velocity. The scale is indicated in physical kpc at the top of each
panel and the dashed circles show the virial radius.

\section{The Ionization Model}
\label{sec:ionmodel}

We developed an ionization model specifically designed for
post-processing application with the hydroART cosmological
simulations.  The details of the code, including comparisons with the
industry standard ionization code Cloudy \citep{ferland98, ferland13}
are presented in \citet{cwc-ioncode}.  The code has also been
successfully applied for observational work \citep[e.g.,][]{ggk-q1317,
cwc-q1317}.  Here, we briefly summarize the ionization modeling of the
gas in the simulations.

For each individual gas cell in the simulation box, the ionization
model calculates the equilibrium ionization state of the gas.  The
ionization model treats photoionization, Auger ionization, direct
collisional ionization, excitation auto-ionization processes, charge
exchange ionization, radiative recombination, dielectronic
recombination, and charge exchange recombination.  If desired, the
effects of each of these processes can be isolated by turning the
process ``off'' or ``on''.  The output of the ionization code is a
simulation box containing the cell equilibrium electron densities, and
the number densities of all ions.  We thus have the ability to study
the spatial distribution of the ions.  If desired, the photoionization
rates and recombination and collisional ionization rate coefficients
in a given cell can be recorded for specified ions.

Three cell properties constrain the gas physics (1) the total hydrogen
density of the cell, $n\subH$, (2) the equilibrium temperature of the
cell, $T$, and (3) the mass fractions of the atomic species in the
cell, which is given as the type II and Ia supernovae yields from
the stellar feedback and metal transport.  Metals up to and including
zinc are incorporated into the ionization model.  The fourth quantity
that governs the gas physics is the spectral energy distribution of
ionizing photons.  The ionization model accounts for the ultraviolet
background (UVB), and/or radiation from the stellar particles
(populations) in the simulated galaxy.

\subsection{Optically Thin Constraint}

A limitation of the current version of the ionization model is that
only optically thin gas can be treated because we presently do not
treat radiative transfer through the grid cells\footnote{We are
  currently developing the techniques to account for shielding
  effects.}.  As discussed in \citet{cwc-ioncode}, we employ the
definition for ``optically thin'' to mean that the optical depth is
less than unity at the hydrogen and helium ionization edges,
which dominate modification of the ionizing SED.  We showed that the
cloud models are constrained to have upper limits on the column
densities of $N\subHI = 1.58\times 10^{17}$ cm$^{-2}$ for neutral
hydrogen, $N\subHeI = 2.77 \times 10^{17}$ cm$^{-2}$ for neutral
helium, and $N\subHeII = 6.34 \times 10^{17}$ cm$^{-2}$ for singly
ionized helium.

Via the relationship $L = N\subXj/f\subXj n\subX$, where $N\subXj$ is
the column density of atomic species X in ionization stage $j$,
$f\subXj$ is the ionization fraction of ion ${\rm X}^{\hbox{\tiny
j}}$, and $n\subX$ is the number density of species X, the upper
limits on column density translate into upper limits on the cell size,
$L_{\rm max}$, for validity of the optically thin constraint.  In
\citet{cwc-ioncode}, we showed that the cell size upper limits can be
written,
\begin{equation}
\begin{array}{rcll}
L_{\rm max}({\HI}) & \simeq & 0.5 \cdot ({0.01}/{f\subHI}) 
({0.01}/{n\subH}) &   \\[3pt]
L_{\rm max}({\HeI}) & \simeq & 9 \cdot
({0.01}/{f\subHeI}) ({0.01}/{n\subH}) & {\rm kpc} \\[3pt]
L_{\rm max}({\HeII}) & \simeq & 20 \cdot ({0.01}/{f\subHeII}) 
({0.01}/{n\subH}) &   \, ,
\end{array}
\label{eq:maxcell}
\end{equation}
assuming a relative abundance of helium to hydrogen of 10\%.

The proper minimum cell size for the simulations at $z=0.54$ (the
redshift of the galaxy we study in this work) is $\simeq 0.03$ kpc.
From Eq.~\ref{eq:maxcell}, we see that only in cases where the product
of the ionization fraction and the hydrogen number density exceed
$10^{-4}$ does the maximum cell size decrease from the fiducial values
of 0.5, 9, and 20 kpc for the respective ionization edges.  In
\citet{cwc-ioncode}, we further showed that the maximum allowed cell
size is below the 30 pc resolution minimum for $\log n\subH \geq -2.2$
when $3 \leq \log T \leq 4$, and for $\log n\subH \geq -1.2$ when
$\log T \simeq 5$.  For $\log T \geq 5.5$, the maximum allowed cell
size is never less than 30 pc.  As such, our ionization model is
currently not entirely valid for ``cold'' cells ($\log T < 4$) with
densities $\log n\subH > -2$ nor for ``warm/hot'' cells ($\log T
\simeq 5$) with densities $\log n\subH > -1$.  The former cells are
found to reside almost exclusively in the ISM of the simulated dwarf
galaxies, and the latter cells are virtually non-existent because the
warm/hot gas is associated with densities in the range $\log n\subH <
-1$.

\subsection{Comparison with Cloudy}

Direct comparisons between our ionization model and Cloudy 13.03
\citep{ferland13} showed that the two codes are in good agreement
\citep{cwc-ioncode}.  The ionization fractions of neutral hydrogen are
virtually identical over the density and temperature ranges $-7 \leq
\log n\subH \leq +1$ and $2 \leq \log T \leq 7$.  The helium
ionization fractions are also in full agreement except for a factor of
2-3 overestimate for neutral helium for $\log T > 5$.  Comparison of
the ionization corrections, $\log f\subXj/f\subHI$, were in agreement
within $\pm 0.05$ for the commonly observed ions Mg$^+$,
C$^{\hbox{\tiny +3}}$, and O$^{\hbox{\tiny +5}}$ over the majority of
the $\log n\subH$--$T$ parameter space.  Most importantly, as we
discuss \citet{cwc-ioncode}, the region of agreement is always within
$\log f\subXj/f\subHI \leq 0.05$ where the ionization fractions of the
target ions are the largest, meaning that dominant ionization stages
where absorption will be most affected are in agreement with Cloudy.
Given that $\log N\subXj \propto \log f\subXj/f\subHI$, we argued that
since typical uncertainties in observed column density measurements
range between $\pm 0.05$ to $\pm 0.1$ in the logarithm, the difference
in the ionization corrections between the two ionization models would
be consistent with typical measurement errors in $\log \NXj$ obtained
from absorption line analysis.

\section{Absorption Line Analysis of the CGM}
\label{sec:simabslines}

\subsection{The Simulation Box and Sightlines}
\label{sec:mklos} 

The redshift of the simulation box is denoted $z_{\rm box}$.  The
position of the center of the galaxy in the box, ${\bf x}_g=(x_{1g} ,
x_{2g}, x_{3g})$, is obtained by locating the center of mass of the
stellar particles surrounding the minimum of the gravitational
potential.  The peculiar velocity of the galaxy in the box is the
velocity of the center of mass of the stellar particles, and is
denoted ${\bf v}_g = (\dot{x}_{1g} , \dot{x}_{2g}, \dot{x}_{3g})$.

\begin{figure*}
\epsscale{0.85}
\plotone{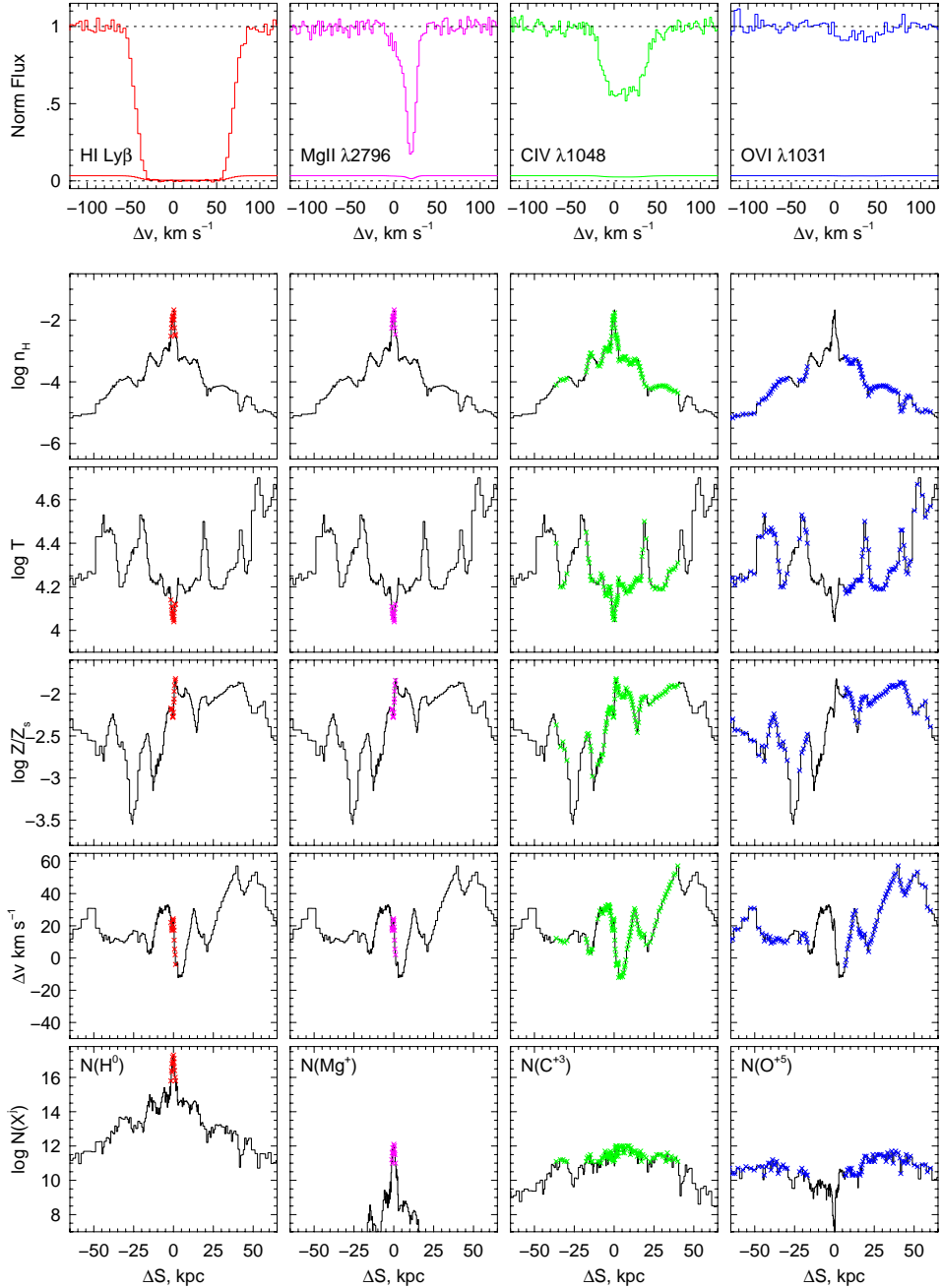}
\caption{Selected ``observed'' absorption profiles and gas cell
  properties for a line of sight through the simulated galaxy (LOS
  0092).  (upper panel) Absorption profiles determined from
  Eqs.~\ref{eq:taui}--\ref{eq:finalcounts} ({\it HST}/COS for
  $SNR=30$) as a function of velocity relative to galaxy systemic, for
  the neutral hydrogen {\Lyb} transition the blue transitions of the
  zero-volt resonant doublets for the Mg$^{+}$, C$^{\hbox{\tiny +3}}$,
  and O$^{\hbox{\tiny +5}}$ ions.  (lower panels, from top to bottom)
  Gas cell hydrogen number density, $\log n\subH $, temperature, $\log
  T$, metallicity, $\log Z/Z_{\odot}$, line of sight velocity, $\Delta
  v$, and ion column density, $\log \NXj$, as a function of line of
  sight position $\Delta S$. The sightline is perpendicular to the
  plane of the sky, and the zero point, $\Delta S = 0$, corresponds to
  the plane of the sky defined by the galaxy center.  The cell column
  densities are determined from the ionization modeling.  Note that
  even though the absorption profiles are closely aligned in velocity
  space, the absorption is arising in physically distinct regions
  along the line of sight.}
\label{fig:los}
\end{figure*}

First, the ionization model is run on the simulation box, from which
the number densities of all ions are determined for each gas cell.  To
generate ``observed'' quasar spectra (see
Section~\ref{sec:mkspectra}), a line of sight (LOS) is passed through
the simulation box from the vantage point of an ``observer'' viewing
the galaxy on the plane of the sky.  Each LOS is defined by (1) an
impact parameter, $D$, (2) a position angle on the plane of the sky,
$\Phi$, which ranges from $\Phi = 0^{\circ}$ to $360^{\circ}$, and (3)
the inclination, $i$, of the simulated galaxy with respect to the LOS
direction.  

The orientation of the galaxy in the simulation box is defined by the
angular momentum vector of the star particles.  Once $D$, $\Phi$, and
$i$ are specified, we determine the direction cosines $(\ell, m, n )$
of the LOS with respect to the box coordinate system.  For an
individual simulated galaxy, we can create and study an arbitrary
number of randomly oriented or parallel LOS.  This formalism allows
the opportunity study the relationship between galaxy orientation and
absorption line properties \citep[e.g.,][]{bordoloi11, bouche12,
  ggk-orientation, mathes14}.

The position along the LOS for cell $i$ is
\begin{equation}
\Delta S_i = \left[ \ell(x_{1i} - x_{1g})^2 + m(x_{2i} -
x_{2g})^2 + n(x_{3i} - x_{3g})^2 \right] ^{1/2} ,
\end{equation}
where ${\bf x}_i = (x_{1i} , x_{2i}, x_{3i})$, measured in
kiloparsecs, is the position of cell $i$ intercepted by the LOS.  The
plane of the sky is defined as the plane perpendicular to the LOS
intersecting $\Delta S = 0$.  Since the LOS unit vector is $\hat{\bf
  s} = \ell\,\hat{\bf i} + m\,\hat{\bf j} + n\,\hat{\bf k} $, the cell
LOS velocity with respect to the simulation box is $v_i = \hat{\bf
  s}\cdot {\bf v}_i$, where ${\bf v}_i = (\dot{x}_{1i}, \dot{x}_{2i},
\dot{x}_{3i})$ is the cell velocity vector, and the observed redshift
of the cell is $z_{i} = z_{\rm box} + (v_i/c)(1+z_{\rm box})$.  The
LOS systemic velocity of the galaxy is $v_g = \hat{\bf s}\cdot {\bf
  v}_g$, and the ``observed'' redshift of the galaxy is $z_{g} =
z_{\rm box} + (v_g/c)(1+z_{\rm box})$.

The column density of ion ${\rm X}^{\hbox{\tiny j}}$ for each cell
along the LOS is $\NiXj = \niXj L_i = \fiXj n\subiX L_i$, the product of
the number density of the ion in the cell and the pathlength of the LOS
through the cell, $L_i$, which is the true length of the LOS vector
through the cell computed as the pathlength from the entry to the exit
points on the cell walls.

For this investigation, we ran 1000 LOS through the simulated galaxy
from the perspective of a face-on orientation.  The LOS range from $0
\leq D \leq 90$ kpc, corresponding to $0 \leq D/R_{\rm vir} \leq 1.5$.
In Figure~\ref{fig:los} (lower set of panels), we illustrate selected
cell physical quantities as a function of the LOS position, $\Delta
S$, over the range $-65 \leq \Delta S \leq 65$ kpc (corresponding to
$R \simeq R_{\rm vir} \simeq 1$) for LOS 0092 through the simulated
galaxy shown in Figure~\ref{fig:simshots}.  This LOS is at $D=11$ kpc,
which corresponds to $D/R_{\rm vir} = 0.2$.  We selected this
particular LOS for illustration purposes because it gives rise to both
low ionization and high ionization absorption, and therefore provides
insights to both gas phases.  Overall, LOS 0092 is typical of the many
LOS that probe the simulated galaxy in the range $0.1 \leq D/R_{\rm
  vir} \leq 0.4$.

From top to bottom in Figure~\ref{fig:los}, we present the gas cell
hydrogen number density, $\log n_i\,\subH $, temperature, $\log T_i$,
gas-phase metallicity\footnote{$Z/Z_{\odot} =
  (x\subM/x\subH)/(x\subM/x\subH)_{\odot}$, where $x\subM$ is the mass
  fraction of all metals and $x\subH$ is the mass fraction of
  hydrogen.}, $\log Z_i/Z_{\odot}$, line of sight velocity, $\Delta
v_i = v_i-v_g$, and ion column density, $\log \NiXj$. The curves are
presented as histograms that show the gas cell pathlengths along the
LOS. Recall that $\Delta S= 0$ is the plane of the sky (defined by the
galaxy center).

Multiple panels of each property are repeated from left to right in
Figure~\ref{fig:los} in order to illustrate which cells contribute to
detected absorption for the {\HI} {\Lyb}, {\MgIIdblt}, {\CIVdblt}, and
{\OVIdblt} transitions (we discuss how these cells are determined in
Section~\ref{sec:isolatingcells}).  The synthetic absorption lines are
shown in the top set of panels (generation of the synthetic spectra is
described in Section~\ref{sec:mkspectra}).  These absorbing cells are
marked with $\times$ points overplotted on the histograms (see
Section~\ref{sec:kin-space} for further discussion).

For this LOS, the CGM of the simulated dwarf has density range $-5
\leq \log n\subH \leq -1.5$, temperature range $4.2 \leq \log T \leq
4.6$, and metallicity range $-3.5 \leq \log Z/Z_{\odot} \leq -2$.
Positive temperature spikes of $\sim 0.3$ dex occur in regions of
$\sim 0.5$--1 dex reductions in density.  From the behavior of the LOS
velocity, there is a clear outflow for $\Delta S > 0$ kpc, with a
velocity inversion in the range $+10 \leq \Delta S > +20$ kpc, which
is characterized by a drop in density and increase in temperature.
Note the $\sim 0.5$ dex drop in metallicity at $\Delta S = +15$ kpc
that proceeds the velocity inversion.  Visual inspection of density,
temperature, metallicity, and velocity slices of the simulated galaxy
shown in Figure~\ref{fig:simshots} clearly show that the CGM of this
galaxy is as highly variable, dynamic, and complex as this single LOS
example indicates.

The computation of the cell column densities (bottom panels of
Figure~\ref{fig:los}) rely on the ionization modeling.  Based upon our
criteria for optically thin gas as determined by the optical depth at
the hydrogen and helium ionization edges, we find that even in the
cells where the H$^{\hbox{\tiny 0}}$ and Mg$^{+}$ column densities are
highly peaked, only one gas cell along the LOS barely violates the
optically thin criteria.  Note that the C$^{\hbox{\tiny +3}}$ and
O$^{\hbox{\tiny +5}}$ column densities have a relatively flat
distribution along the LOS out to $\Delta S \sim \pm 50$ kpc.

\subsection{Generation of Simulated Spectra}
\label{sec:mkspectra}

\begin{figure*}
\epsscale{1.00}
\plotone{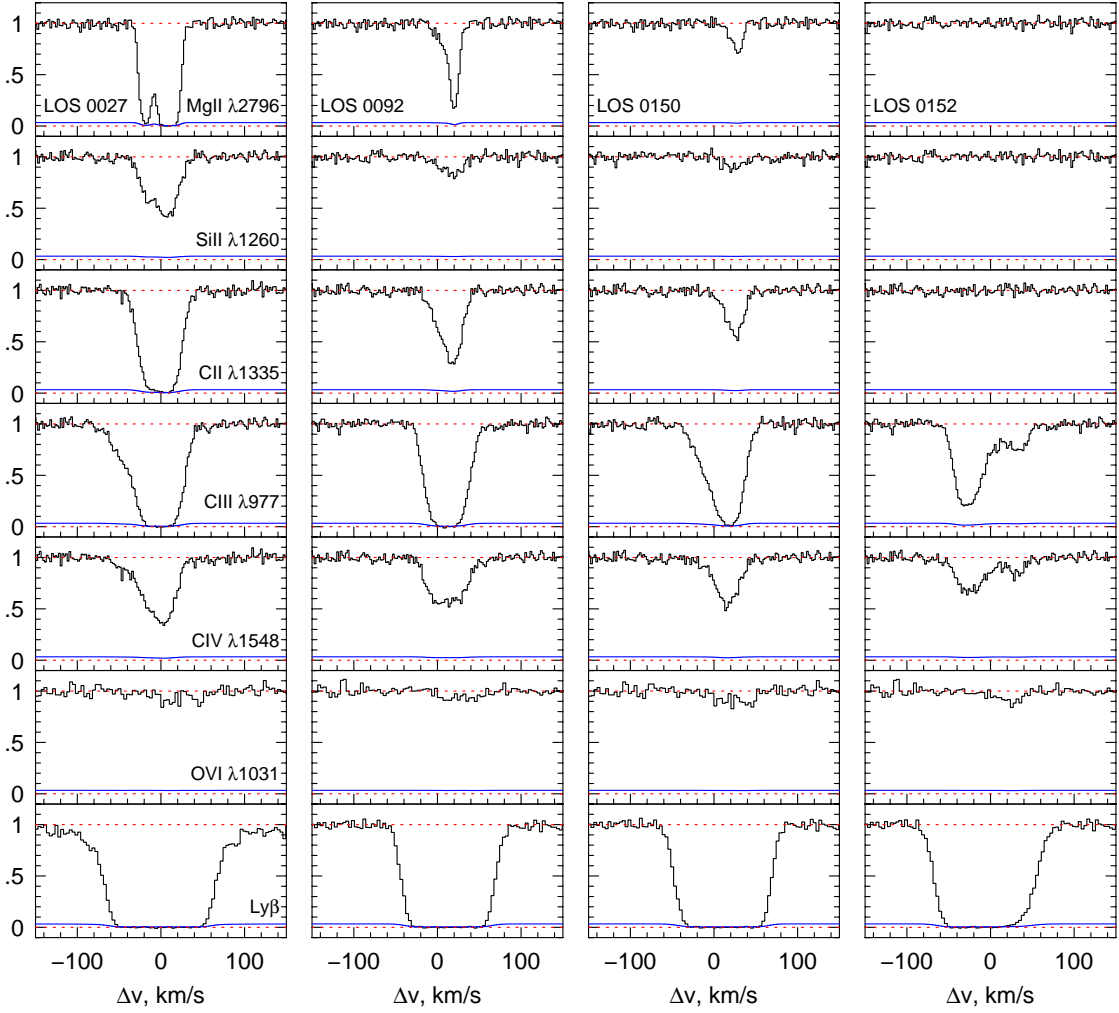}
\caption{Selected synthetic absorption profiles for four LOS through
  the simulated galaxy.  From left to right, the LOS are 0027, 0092,
  0150, and 0152, with increasing impact parameter range from $D\simeq
  9$ to 30~kpc ($D/R_{\rm vir} \simeq 0.15$ to 0.5).  The profiles and
  LOS properties for LOS 0092 (second from the left) have been further
  illustrated in Figure~\ref{fig:los}.  The spectra have resolution
  have $SNR=30$, and correspond to the HIRES and COS FUV and NUV
  gratings.}
\label{fig:profiles}
\end{figure*}

The simulated absorption spectra are generated with our code Specsynth
with the assumption that each cell, $i$, contributes to the optical
depth as if it were an isothermal ``cloud''.  For a given transition
from ion ${\rm X}^{\hbox{\tiny j}}$, the optical depth as a function
of observed wavelength is computed from
\begin{equation}
\tau_i (\lambda) = \NiXj
 \frac{\sqrt{\pi}e^2}{m_e c^2} \frac{f\lambda^2_0}{\Delta
  \lambda _i} \, U (a,b) ,
\label{eq:taui}
\end{equation}
where the physical constants have their usual meaning, $\lambda_0$ is
the rest-frame wavelength of the transition, $f$ is the transition
oscillator strength, and $\NiXj$ is the column density of the ion.
The term
\begin{equation}
\Delta \lambda _i = \frac{\lambda _0}{c} \left\{ \frac{2kT_i}{m\subX} \right\} ^{1/2}
\end{equation}
is the Doppler width, where $m_{_{\rm X}}$ is the mass of the
ionic species.  The Voigt function, $U(a,b)$, is computed as the real
part of the complex probability function using the code Cpf12
\citep{humlicek79} with unitless parameters
\begin{equation}
\begin{array}{rl}
a = \displaystyle \frac{1}{\Delta \lambda _i} 
       \frac{\lambda - \lambda_c}{(1+z_g)} \, , & \,\,
b = \displaystyle \frac{1}{\Delta \lambda _i} 
       \frac{\Gamma \lambda^2_0}{4\pi c} \, ,
\end{array}
\end{equation}
where the factor $1+z_g$ ensures that the wavelength difference,
$\lambda - \lambda _c$, is co--moving corrected, and where $\lambda _c
= \lambda_0 (1+z_i)$ is the redshifted central wavelength of the
absorbing cell.  The resulting normalized counts in the spectrum prior
to being recorded by an instrument are obtained by
\begin{equation}
{\cal I}'(\lambda) = \prod _{i=1}^{N_c} \exp [-\tau_i (\lambda)] \, ,
\end{equation}
where $N_c$ is the number of cells along the LOS.  

The spectrum is then ``passed through'' an instrument.  The choice of
instrument would be dictated by which ions and redshifts are being
studied so that the spectra can be directly comparable to
observational data.  For example, {\HI} Lyman series lines and the
{\OVI} doublet at low redshifts, i.e., $z<0.3$, are observed in COS
G130M spectra, whereas the {\MgII} doublet at $z>0.3$ is studied in
Keck/LRIS, Keck/HIRES and/or VLT/UVES spectra.  Having chosen an
instrument, we first convolve ${\cal I}'(\lambda)$ with the instrument
spread function (ISF), $\Phi (\lambda ' - \lambda)$, yielding the
instrument convolved normalized spectrum,
\begin{equation}
{\cal I}(\lambda) = \Phi (\lambda ' - \lambda) \ast {\cal I}'(\lambda ') \, .
\end{equation}

The convolved normalized spectrum is then sampled with the pixelization
$\Delta \lambda _{\rm pix}$, of the chosen instrument.  Finally,
Gaussian deviate noise is added on a pixel by pixel basis assuming a
fixed signal-to-noise ratio per pixel, $SNR$, and adopting the
instrumental read noise, ${\rm RN}$.  The read noise is applied in
units of electrons, and not in digital number (which is smaller by the
readout amplifier gain factor) because properly modeling the Poisson
statistics requires electron counts.  The adopted $SNR$ ideally should
be selected to reflect the average $SNR$ of the observed spectra
comprising surveys targeting the transition being studied in the
simulations.  In this way, the detection sensitivities of
observational surveys are emulated for a direct comparison between
spectra from simulations and real world spectra.

The normalized uncertainty spectrum due to Poisson statistics is given
by
\begin{equation}
\sigma_{\cal I}(\lambda) = \displaystyle \frac{ \left[ {\cal I}_c
    {\cal I}(\lambda) + {\rm RN}^2 \right] ^{1/2} } {{\cal I}_c} \, ,
\end{equation}
where 
\begin{equation}
{\cal I}_c = \displaystyle \frac{(SNR)^2}{2} \left[ 1 + \sqrt{ 1 + 4 \left(
\frac{\rm RN}{SNR} \right) ^2 } \right]
\end{equation}
approximates the continuum counts for the desired $SNR$
\citep{cwcthesis}.  To account for the additional uncertainty due to
placement of the continuum fit (as required for observational
spectra), we adopt an approximation that the uncertainty due to
continuum placement is proportional to the Poissonian uncertainty in
the continuum, i.e., $h\, \sigma_{{\cal I}_c}(\lambda)$, where ${\cal
I}(\lambda)=1$ \citep[see][]{ss92}.  The final normalized uncertainty
spectrum is then
\begin{equation}
\begin{array}{rcl}
\sigma (\lambda) \!\!\!\! & = & \!\!\!\!
\sqrt{ \sigma^2_{\cal I}(\lambda) +  
 h^2 \sigma^2_{{\cal I}_c}(\lambda) } \\[10pt]
\!\!\!\! & = & \!\!\!\! \displaystyle
\frac{\displaystyle
\sqrt{ {\cal I}_c {\cal I}(\lambda) + {\rm RN}^2 +
h^2\left[ {\cal I}_c + {\rm RN}^2 \right] } }
{\displaystyle {\cal I}_c} \, ,
\end{array}
\end{equation}
To determine the proportionality constant, $h$, we undertook a blind
experiment.  We generated 100 synthetic spectra from several common
instruments with a range of $SNR$ (from 5 to 50) and added continuum
shapes ranging from from 2nd order to 7th order Legendre polynomials.
Each spectrum had 2048 pixels.  We then interactively continuum fit
the spectra and computed the mean standard deviation (over all pixels)
between the noiseless input continua and the blindly fitted smooth
continua.  This yielded $h = 0.4$, which, in continuum regions,
corresponds to an 8\% increase for the final uncertainty
relative to the Poisson-only uncertainty.

The final pixelated normalized spectrum with noise is computed from
\begin{equation}
I(\lambda) = {\cal I}(\lambda) + G_{\rm pix} \sigma(\lambda) ,
\label{eq:finalcounts}
\end{equation}
where $G_{\rm pix}$ is a random unit Gaussian deviate generated on a
per pixel basis.

For each LOS (and there can be an arbitrary number per simulated
galaxy), an individual synthetic spectrum and uncertainty spectrum is
generated for each ion/transition that we aim to study.  The observed
wavelength range of a given synthetic spectrum is set such that the
spectrum covers $\Delta v = \pm 1000$~{\kms} with respect to the
system velocity of the simulated galaxy.  However, this range can be
adjusted and redefined with ease.

\begin{figure*}
\epsscale{1.10}
\plotone{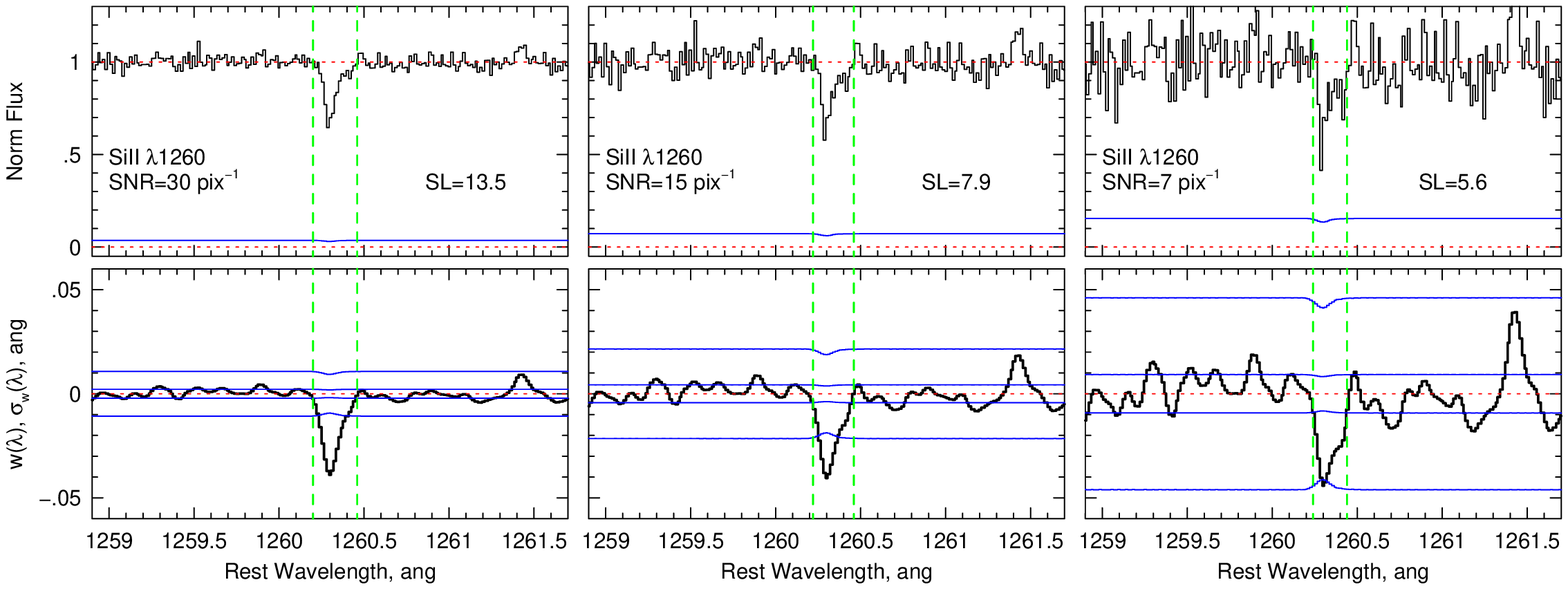}
\caption{Detection method and criteria for objectively defining
  absorption features. (top panels) De-redshifted {\SiII} $\lambda
  1260$ absorption (COS G185M/1621B) for three $SNR$ values (left)
  $SNR=30$, (center) $SNR=15$, (right) $SNR=7$ per pixel.  The blue
  spectrum provides the $1~\sigma$ uncertainty in the normalized flux.
  (bottom panels) The rest-frame equivalent width per resolution
  element, $w(\lambda)$.  The bracketing blue spectra are the $\pm
  1\sigma_w(\lambda)$ and $\pm 5\sigma_w(\lambda)$ equivalent width
  detection thresholds for unresolved absorption lines, where
  $\sigma_w(\lambda)$ is the equivalent width uncertainty spectrum.  An
  absorption feature is detected at wavelength $\lambda$ when
  $w(\lambda) \leq -5\sigma_w(\lambda)$.  The spectral range of the
  absorption feature (green vertical lines) is defined at the
  wavelengths where $w(\lambda) \leq -\sigma_w$ is first recovered,
  i.e., $w(\lambda)$ becomes consistent with $\sigma_w(\lambda)$.  For
  each realization of the absorption line , the detection significance
  level, $SL$ is provided, where $SL=W_r/\sigma_{_{W_r}}$.}
\label{fig:sysanal}
\end{figure*}

In Figure~\ref{fig:profiles}, we present example synthetic spectra for
four LOS (0027, 0092, 0150, and 0152, from left to right in order of
increasing impact parameter) through the the simulated galaxy shown in
Figure~\ref{fig:simshots}.  The impact parameter range from $D\simeq
9$ to 30~kpc ($D/R_{\rm vir} \simeq 0.15$ to 0.5).  From top to
bottom, we show the the commonly studied {\MgII}~$\lambda 2796$,
{\SiII}~$\lambda 1260$, {\CII}~$\lambda 1335$, {\CIII}~$\lambda 977$,
{\CIV}~$\lambda 1548$, {\OVI}~$\lambda 1031$ transitions, and the
{\HI} {\Lyb} transition.  For this example, we adopt $SNR=30$ per
pixel and show only the ``blue'' member of the doublets.

Since the simulated galaxy is at $z=0.54$, the {\MgII} lines are
redshifted into the optical at 4306~{\AA} so we adopt the HIRES
instrument ($R=45,000$) for this doublet.  We model the HIRES ISF as a
unit-area Gaussian with $\sigma = c/2.35R = 2.83$~{\kms},
corresponding to 3 pixels per resolution element ($6.6$~{\kms}).  The
{\Lyb}, {\OVI} and {\CIII} lines are observed in the FUV G160M grating
of COS ($R \simeq 20,000$), the {\CII} and {\CIV} lines are observed
in the NUV G225M grating of COS ($R\simeq 22,000$), and the {\SiII}
line is observed with the NUV G185M grating of COS ($R \simeq
18,000$).  For the COS ISFs we employ the on-line Lifetime Position 2
tabulated theoretical line spread functions \citep{kriss11}.  We
determine the ISF at a given observed wavelength using cubic spline
interpolation.

LOS 0092 (second from left) is the LOS illustrated in
Figure~\ref{fig:los}.  Note that as impact parameter is increased the
low ionization absorption diminishes and the {\CIII} and {\CIV}
absorption develops somewhat greater kinematic complexity.  We further
discuss the nature of the absorption for LOS 0092 in
Section~\ref{sec:discuss}.

\subsection{Analysis of Simulated Spectra}
\label{sec:sysanal}

In order to emulate observational analysis common to quasar absorption
line studies, we adopt objective methods used (or that should be used)
for observed spectra.  We use a fully automated version of our
graphical interactive code Sysanal, which we have applied for several
other works \citep[e.g.,][]{cv01, evans-phd, cwc-q1317, ggk-q1317,
  evans13, mathes14}.  After all measured quantities are calculated,
the final step of the process is the generation of a single table for
each ion that contains the absorption properties for all LOS.  These
tables can then be analyzed to study the CGM properties of the
simulated galaxies.

\subsubsection{Objective Absorption Line Detection}

To locate statistically significant absorption features in the
synthetic spectra, we employ the methods described in \citet{weakI}
and \citet{archiveI}, which are derived from \citet{schneider93}.

First, the spectrum is converted to an equivalent width spectrum,
$w(\lambda)$, and the uncertainty spectrum is converted to an
equivalent width uncertainty spectrum, $\sigma_w(\lambda)$.  The
equivalent width spectrum provides the observed (not rest-frame)
equivalent width per resolution element for an unresolved absorption
feature as a function of observed wavelength and the equivalent width
uncertainty spectrum provides the $1~\sigma$ observed equivalent width
detection threshold as a function of observed wavelength.  The
computation of $w(\lambda)$ and $\sigma_w(\lambda)$ requires
convolving the normalized and pixelized ISF with the flux decrements
in each pixel of the spectrum \citep[see Section 3.1.1 and Equations 1
  and 2 of][]{archiveI}.

An absorption feature is objectively defined in spectral regions where
$w(\lambda) \leq -N\sigma_w(\lambda)$, where we adopt $N=5$ for
singlet absorption lines.  For doublets, we adopt $N=5$ for the blue
(higher oscillator strength) member of a doublet and $N=3$ for the red
member of a doublet.  For a doublet to be ``detected'', both members
must satisfy the above detection criteria.  However, to facilitate
direct comparison with observational data from various surveys that
may adopt different criteria, the detection criteria can easily be
tailored to those of any survey.

The lower and upper wavelength limits over which the absorption
feature is defined are taken to be where $w(\lambda) \geq
-1\sigma_w(\lambda)$ is first recovered while scanning the spectrum
blueward and then redward of the wavelength first satisfying the
detection criterion.  Using this method, several ``sub-features'',
absorption separated by continuum, can be uniquely defined and
analyzed \citep[see Figure~1 of][]{cv01}.

In Figure~\ref{fig:sysanal}, we illustrate the detection method for
objectively defining absorption features.  In the upper panels, we
present the rest-frame {\SiII} $\lambda 1260$ synthetic absorption
lines for the COS G185M/1921 grating for Stripe B.  We vary the $SNR$
values (left) $SNR=30$, (center) $SNR=15$, (right) $SNR=7$ per pixel.
The blue spectrum provides the $1~\sigma$ uncertainty in the
normalized flux.  In the lower panels of Figure~\ref{fig:sysanal}, we
present the rest-frame equivalent width spectrum, $w(\lambda)$.  We
also present the $\pm 1\sigma_w(\lambda)$ and $\pm 5\sigma_w(\lambda)$
rest-frame equivalent width detection thresholds for unresolved
absorption lines as the bracketing blue spectra.  Note that the
$5~\sigma$ detection threshold changes from $\simeq 0.011$~{\AA} for
$SNR=30$ to $\simeq 0.045$~{\AA} for $SNR=7$.  For $SNR=30$ and 15,
several pixels across the profiles are clearly significant at greater
than the $5~\sigma$ level.  However, for $SNR=7$, the absorption
feature is on the verge of detection; had the noise characteristics
been different, this feature may not have been formally detected.  The
green vertical lines mark the wavelength range over which the
absorption profiles are defined for the purpose of quantifying the
absorption.

Sysanal works simultaneously on all ion/transitions for a given LOS.
Once all synthetic spectra for a LOS are objectively searched for
absorption features, and the spectral ranges of the detected
absorption features are determined, the code then computes the
observed and rest-frame equivalent widths, doublet ratios (when
applicable), the flux decrement weighted velocity centers, velocity
widths, and velocity asymmetries.  Formal uncertainties in these
quantities are also computed.

In the case of nondetections, the $3~\sigma$ upper limits are computed
for the observed and rest-frame equivalent widths.  In the case where
multiple sub-features are detected, they are individually measured in
addition to the measurement of the ``total system'' quantities (all
sub-features treated as a single absorption feature).  The
mathematical computation of these quantities, originally based on the
work of \citet{ss92}, is given in \citet{cwcthesis} and in the
Appendix of \cite{cv01}.  Additional details are provided in
Sections 3.1--3.4 of \citet{evans-phd}.

\subsubsection{Apparent Optical Depth Spectra}

Sysanal also creates an apparent optical depth (AOD) column density
spectrum and uncertainty AOD spectrum for each ion/transition
\citep{savage91}.  From these spectra, we determine the ``best'' AOD
column density spectrum for a given ion.  For ions with multiple
transitions, such as {\HI} and the common metal-line doublets, we
employ the following procedure.  For a given velocity pixel, we take
the optimal weighted mean AOD column density of the transitions if (1)
more than one transition for a given ion has measured values (not
lower or upper limits) of the AOD column density, and (2) unresolved
saturation \citep[see][]{savage91} is not present.  The uncertainties
are propagated in the standard fashion.  If all but one transition
yields a limit, the transitions with measured values are adopted.  In
the case of unresolved saturation, the adopted AOD column density is
taken to be that with the highest oscillator strength.  In the case of
upper limits in all transitions, the highest oscillator strength
transition is adopted, and in the case of lower limits in all
transitions, the lowest oscillator strength transition is adopted.
This process yields the AOD column density profile for a given ion,
$N_a({\rm X}\subj;\Delta v)$ [atoms cm$^{-2}$ ({\kms})$^{-1}$], from
which we compute the integrated AOD column densities, $N_a({\rm
  X}\supj)$ for each ion,
\begin{equation}
N_a({\rm X}\supj) = \int _{\Delta v^{(-)}}^{\Delta v^{(+)}} \!\!\!\! 
N_a({\rm X}\supj;\Delta v)\, d(\Delta v) \quad \hbox{atoms cm$^{-2}$} \, ,
\label{eq:Naod}
\end{equation}
where the integration is over the spectral region over which the
absorption profile is detected (see the vertical green lines in
Figure~\ref{fig:sysanal}).  If saturation persists over a minimum of
three adjacent pixels, i.e., $N_a({\rm X}\supj;\Delta v)$ is a lower
limit over an extended velocity range approaching a resolution
element, we quote a $3~\sigma$ lower limit for $N_a({\rm X}\supj)$.
We discuss comparison of the AOD spectra to the simulation gas
properties in Section~\ref{sec:kin-space}.

\subsubsection{Voigt Profile Decomposition}

The final spectral analysis step is VP decomposition, which yields the
column densities, $N$, Doppler $b$ parameters, and observed redshifts,
$z$, of multiple VP components.  To obtain an initial model for each
ion, we run Autovp\footnote{Autovp was originally written R. Dav\'{e}.
  We modified the program to incorporate convolution with the ISF.}
on the spectra.  Redshifted transitions from different ions can fall
in wavelength ranges appropriate for different spectrograph/grating
combinations (i.e., COS, HIRES, LRIS, etc.), so that different
transitions from a single LOS are likely to be measured with different
resolutions.  We properly account for the ISF appropriate for the
instrument with which a given synthetic absorption line was created.

Due to the different kinematics of the lower and higher ionization
gas, we have found that fully automating the VP decomposition of the
synthetic spectra to be challenging and we are still developing our
approach.  Currently, we run Autovp on the lower ionization profiles
and select the profile with the largest number of VP components as the
initial kinematic template for the low ionization transitions.  We
repeat the process for the higher ionization transitions.

To obtain the final model, we then run our minimization code Minfit
\citep{cwcthesis, cv01, cvc03}.  Minfit refines the model using the
maximum likelihood modification of the Levenberg-Marquardt algorithm
Dnls1 \citep{more78}, which minimizes the sum of the squares of $M$
nonlinear functions in $N$ variables\footnote{Dnls1 is part of the
Slatec package publicly available at the Netlib Repository ({\it
http://www.netlib.org\/}) sponsored by The University of Tennessee and
the Oak Ridge National Laboratory.}.  We adopt the model with the
fewest VP components that are statistically significant at the 97\%
confidence level by applying an $F$-test on the $\chi^2$
distribution.  Components with the largest fraction errors,
\begin{equation}
f_{\rm err} = \sqrt{
\left(\frac{dN}{N}\right)^2 +
\left(\frac{db}{b}\right)^2 +
\left(\frac{dz}{z}\right)^2 
} \, ,
\end{equation}
are tested for significance in descending order of $f_{\rm err}$.  If
a component is not significant, it is removed from the model, and the
process is repeated until all components are significant.  

Full details of the most up-to-date version of Minfit and the fitting
process are described in \citet{evans-phd}.  The difference for our
application is that we adopt a unique VP model for the lower
ionization gas and for the higher ionization gas.  Though not perfect
as a description of the complex multiphase gas structures that give
rise to the absorption lines, it does provide us a formalism for
segregating the absorption into two gas phases.  Further development
is under way.

\begin{figure}
\epsscale{0.925}
\plotone{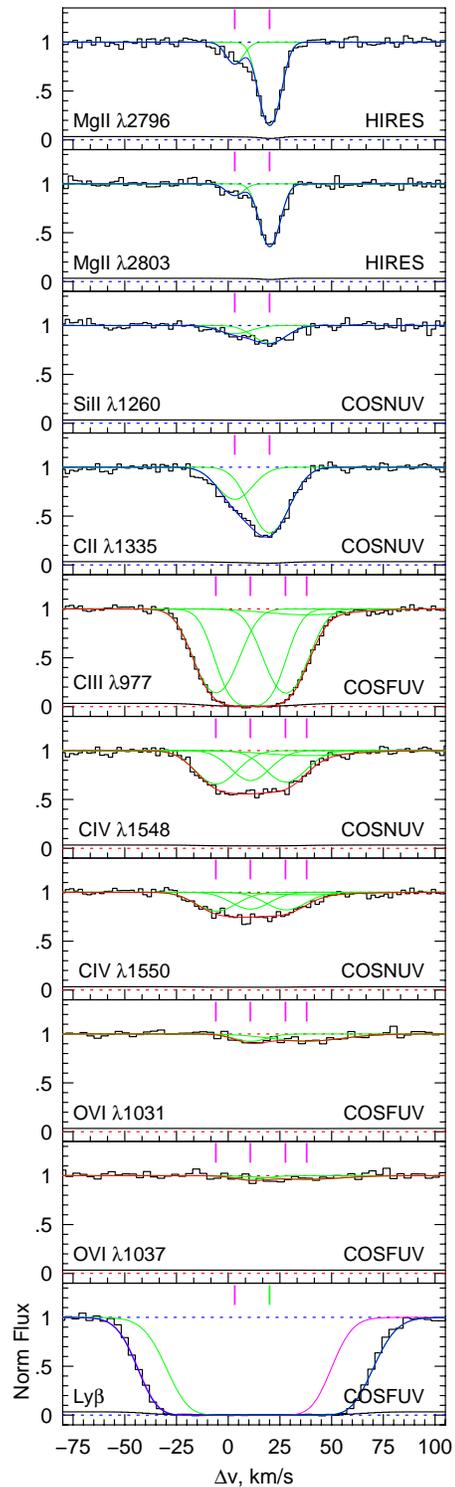}
\caption{Voigt profile decomposition of selected transitions for the
  absorption profiles for LOS 0092 shown in Figure~\ref{fig:los}. The
  lower-ionization transitions of {\MgII}, {\SiII}, {\CII}, and {\HI}
  (blue curves) were fitted simultaneously and separate to the
  higher-ionization transitions of {\CIII}, {\CIV}, and {\OVI} (red
  curves).  Magenta and green curves provide the individual compoents.
  Though component velocities are tied between ions, the column
  densities and allowed to vary freely.  The $b$ parameters of each
  component are tied between ions through thermal scaling. The ticks
  above the continua provide the number of components and their
  velocity centroids.  The instruments assumed for generation of the
  synthetic spectra and therefore the ISFs used in the VP
  decomposition are given in the lower right of each panel.}
\label{fig:vpfit}
\end{figure}

In Figure~\ref{fig:vpfit}, we show an example of VP decomposition for
LOS 0092.  This is the LOS illustrated in Figure~\ref{fig:los}.  For
this LOS, the lower ionization species are taken to be {\MgII},
{\SiII}, {\CII}, and {\HI} (shown as the blue fits), whereas the
higher ionization species are taken to be {\CIII}, {\CIV}, and {\OVI}
(shown as the red fits).  The individual VP components are shown as
magenta (except for {\HI} {\Lyb}, where, for clarity, one component is
magenta and the second is green). The LOS velocities of the low
ionization VP components are tied together and all lower ionization
transitions are fitted simultaneously.  The same holds for the higher
ionization transitions.  We assume purely thermal line broadening for
each VP component.  In this mode, the component temperatures are the
actual fitting parameters, from which the $b$ parameters are computed
from $b\subXj = \sqrt{2kT/m\subX}$ for ion X$\supj$, where $n\subX$ is
the mass of ion X.

For reference, we list the VP component parameters in
Table~\ref{tab:vpfit}.  Presented are (1) the ion and transition
fitted, (2) the LOS velocity, $\Delta v$, (3) the column density,
$\log N$, (4) the Doppler $b$ parameter, and (5) the temperature.
Upper limits are quoted at the $3~\sigma$ level.  We discuss
comparison of the VP fit parameters to the simulation gas properties
along this LOS in Section~\ref{sec:kin-space}.

\begin{deluxetable}{lrrrr}
\tablecolumns{5}
\tablewidth{0pt}
\setlength{\tabcolsep}{0.06in}
\tablecaption{VP Decomposition of LOS 0092\label{tab:vpfit}}
\tablehead{
  \colhead{(1)}             &
  \colhead{(2)\tablenotemark{a}}  &
  \colhead{(3)\tablenotemark{b}} &
  \colhead{(4)\tablenotemark{c}} &
  \colhead{(5)}             \\[2pt]
  \colhead{Ion/}             &
  \colhead{$\Delta v$}             &
  \colhead{$\log N$}             &
  \colhead{$b$}             &
  \colhead{$\log T$}             \\
  \colhead{Transition}      &
  \colhead{({\kms})}        &
  \colhead{(cm$^{-2}$)}     &
  \colhead{({\kms})}        &
  \colhead{(K)}             }
\startdata
\cutinhead{Low Ionization Phase} \\[-3pt]
{\HI}                       &  3.20  & $16.27\pm0.15$ & $20.80\pm0.85$ & $4.42\pm0.04$ \\ 
                            &  20.01 & $16.04\pm0.06$ & $23.92\pm0.46$ & $4.54\pm0.02$ \\[3pt]
{\MgII}                     &  3.20  & $11.77\pm0.03$ & $ 4.24\pm0.17$ & $4.42\pm0.03$ \\ 
$\lambda\lambda 2796,2803$  &  20.01 & $12.73\pm0.01$ & $ 4.87\pm0.09$ & $4.54\pm0.02$ \\[3pt]
{\SiII}                     &  3.20  & $11.72\pm0.07$ & $ 3.94\pm0.16$ & $4.42\pm0.04$ \\ 
$\lambda 1260$              &  20.01 & $12.10\pm0.04$ & $ 4.53\pm0.09$ & $4.54\pm0.02$ \\[3pt]
{\CII}                      &   3.20 & $13.31\pm0.02$ & $ 6.03\pm0.25$ & $4.42\pm0.04$ \\ 
$\lambda 1335$              &  20.01 & $13.86\pm0.02$ & $ 6.93\pm0.13$ & $4.54\pm0.02$ \\
\cutinhead{High Ionization Phase} \\[-3pt]
{\CIII}                     & $-5.89$ & $13.41\pm0.15$ & $ 9.33\pm0.76$ & $4.80\pm0.04$ \\ 
$\lambda 977$               &  10.69  & $14.66\pm1.42$ & $ 8.09\pm4.89$ & $4.68\pm0.26$ \\
                            &  27.80  & $13.42\pm0.18$ & $ 9.01\pm1.05$ & $4.77\pm0.05$ \\
                            &  38.05  & $12.19\pm0.62$ & $26.33\pm8.60$ & $5.70\pm0.31$ \\[3pt]
{\CIV}                      & $-5.89$ & $13.13\pm0.09$ & $ 9.33\pm0.76$ & $4.80\pm0.04$ \\ 
$\lambda\lambda 1548,1550$  &  10.69  & $13.05\pm0.23$ & $ 8.09\pm4.89$ & $4.68\pm0.26$ \\
                            &  27.80  & $13.09\pm0.11$ & $ 9.01\pm1.05$ & $4.77\pm0.05$ \\
                            &  38.05  & $12.52\pm0.61$ & $26.33\pm8.60$ & $5.70\pm0.31$ \\[3pt]
{\OVI}                      & $-5.89$ &     $<11.36$   & $\cdots$       & $\cdots$      \\ 
$\lambda\lambda 1031,1027$  &  10.69  & $12.62\pm0.56$ & $ 7.01\pm4.24$ & $4.68\pm0.26$ \\
                            &  27.80  &     $<11.31$   & $\cdots$       & $\cdots$      \\
                            &  38.05  & $12.95\pm0.63$ & $22.81\pm6.12$ & $5.70\pm0.31$ 
\enddata
\tablenotetext{a}{The velocities of the components are tied together for each phase.}
\tablenotetext{b}{The column densities vary freely from ion to ion.}
\tablenotetext{c}{The Doppler parameters are tied via thermal broadening for each phase.}
\end{deluxetable}

\subsection{Locating ``Absorbing'' Cells}
\label{sec:isolatingcells}

One of our aims is to develop methods to directly compare the ``true''
properties of the simulated CGM to those inferred from observations.
Comparisons of the simulated CGM and the observed CGM must account for
the gas being probed by the absorption lines.  Isolating the gas 
that is responsible for detected absorption allows for {\it direct\/}
comparison between the measured ``observed'' quantities from synthetic
absorption line analysis and the physical properties of the gas.  As
such, it is centrally important to identify which grid cells along a
given LOS are ``detected'' in simulated absorption lines.

In order to isolated the cells that contribute to the absorption
profile from a given ion, we adopt a differencing technique.  For a
given ion, the cells with LOS velocities aligned within the
objectively defined velocity range of the absorption profiles are
sorted into descending column density order.  An identification number
for each cell from the simulation box is included in the list.  For a
given ion, we then iteratively regenerate synthetic spectra for the
transition with the largest oscillator strength by progressively
omitting one gas cell at a time until the equivalent width of the
profile stops changing by a defined percent difference.  The spectra
for this exercise are appropriately convolved with the instrumental
spread function and pixelated according to the selected spectrograph
and grating settings; however, the test spectra are noiseless.  After
testing various percent differences, we adopted 5\% difference.

To elaborate, we start with the highest column density cell in the
velocity range of the absorption profile, remove the cell, and
recompute the profile.  If the equivalent width is reduced by more
than a 5\% difference, $\left| \Delta W_r/W_r \right| \geq 0.05$, we
consider this cell to contribute significantly to the absorption.  We
then advance to the next highest column density cell, regenerate the
profile, and determine if the equivalent width has decreased by more
than 5\% difference.  We repeat the process on successively smaller
column density cells until the change in the equivalent width is less
than a 5\% difference.  The result is that we identify all cells that
contribute more than a 5\% difference to the equivalent width of the
profile. Cells that reside in the velocity range of the absorption
deemed to not contribute to absorption typically have column densities
a factor of 50 to 100 below the cell contributing the highest column
density.

Referring back to Figure~\ref{fig:los}, we overplotted colored data
points showing the properties of the absorption selected gas as a
function of LOS position, $\Delta S$.  As can be seen, the {\HI} and
{\MgII} absorption lines arise in many of the same cells within
close proximity of $\Delta S \sim 0$.  The {\CIV} absorption also
arises in these cells; but the vast majority of the {\CIV}
absorbing cells are distributed over $\Delta S = \pm 40$ kpc, a region
over which the gas phases change substantially along the LOS with a 2
dex variation in $n\subH$ and 0.4 dex variation in $T$.  Note the 
cells at $\Delta S \simeq -25$~kpc, where $\log Z/Z_{\odot}$ dips
below $-3$, do not contribute to {\CIV} absorption (nor {\OVI}
absorption).  Many of the cells that give rise to {\CIV}
absorption also give rise to {\OVI} absorption, however, the physical
extent of {\OVI} absorbing cells along the LOS is greater than that of
the {\CIV} absorbing cells.

\section{Discussion}
\label{sec:discuss}

\begin{figure*}
\epsscale{1.0}
\plotone{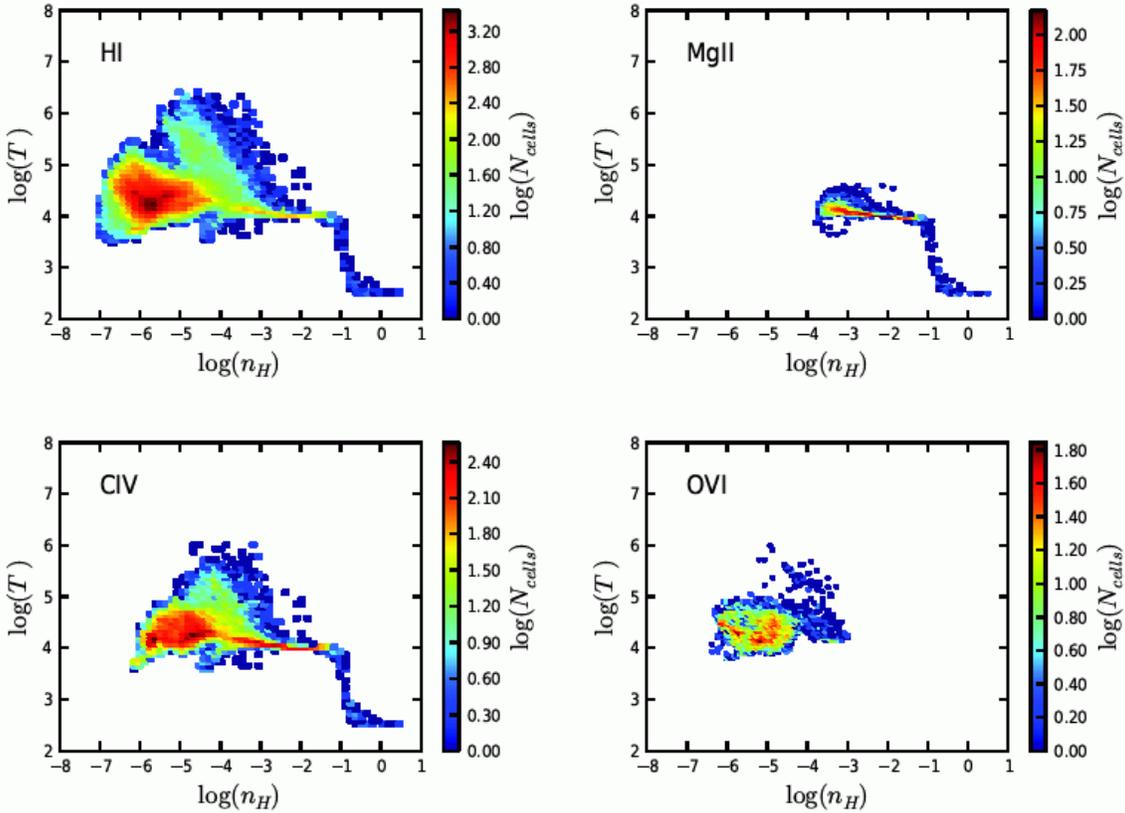}
\caption{The phase diagrams (hydrogen density, $n\subH$, versus
  temperature, $T$) showing the relative number of cells giving
  rise to detected absorption in the synthetic spectra for the four
  ions, {\HI}, {\MgII}, {\CIV}, and {\OVI}.  These cells were selected
  using the criteria described in Section~\ref{sec:isolatingcells} for
  1000 random LOS distributed within $D \leq 1.5 R_{\rm vir}$ through
  the simulated galaxy shown in Figure~\ref{fig:simshots}.  The
  synthetic spectra have the characteristics of the appropriate COS
  NUV or FUV high resolution gratings for a $z\simeq 0.5$ absorption
  line system.  The signal-to-noise ratio is 30 per pixel.  We applied
  a $5~\sigma$ equivalent width detection threshold of $\simeq
  0.1$~{\AA} to these spectra.  Cells with gas phases in the range
  $\log n\subH \geq -1$ and $\log T \leq 3.8$ are from the galaxy ISM;
  the remaining cells are from the galaxy CGM.}
\label{fig:phases}
\end{figure*}

Following the methods described in Sections~\ref{sec:mklos} and
\ref{sec:mkspectra}, we ran 1000 random LOS through the simulated
galaxy (face-on orientation) and generated synthetic absorption line
spectra of {\HI} {\Lya} and {\Lyb}, {\SiII} $\lambda \lambda 1190,
1193$, {\SiII} $\lambda 1260$, {\CII} $\lambda 1036$ and $\lambda
1335$, {\MgIIdblt}, {\CIVdblt}, and {\OVIdblt} transitions.  The
synthetic spectra have $SNR=30$ and the characteristics of the HIRES
instrument and the appropriate COS NUV or FUV high resolution gratings
for a $z\simeq 0.54$ absorption line system.  The $5~\sigma$
rest-frame detection threshold of these spectra is $\sim 0.01$~{\AA}
(see Figure~\ref{fig:sysanal}).  The impact parameters of the LOS
cover the range $0 \leq D/R_{\rm vir} \leq 1.5$, corresponding to
random sky coverage within a projected separation from the galaxy of
$D=90$ kpc.

We then analyzed the absorption line spectra, measuring their
equivalent widths, velocity widths, and column densities, etc., as
described in Section~\ref{sec:sysanal}.  Finally, we determined which
cells contributed to absorption following the methods described in
Section~\ref{sec:isolatingcells}.

\subsection{Phase Space of the CGM}
\label{sec:phases}

In Figure~\ref{fig:phases}, we plot the phase diagrams (hydrogen
density, $n\subH$, versus temperature, $T$) showing the relative
number of cells giving rise to detected absorption in the
synthetic spectra for the four ions, H$^{\hbox{\tiny 0}}$ ({\HI}),
Mg$^{+}$ ({\MgII}), C$^{\hbox{\tiny +3}}$ ({\CIV}), and
O$^{\hbox{\tiny +5}}$ ({\OVI}). For these phase diagrams, we include
only cells that partake in absorption lines having equivalent widths
greater than $0.1$~{\AA}.  This threshold is typical of the
sensitivity threshold of the COS-Halos and COS-Dwarfs surveys
\citep{tumlinson11, tumlinson13, bordoloi14}, though some of the
sightlines in the surveys have deeper detection thresholds.

For {\HI}, {\MgII}, and {\CIV}, the absorbing cells with phases in the
range $\log n\subH \geq -1$ and $\log T \leq 3.8$ are located in the
galaxy ISM (roughly within $D/R_{\rm vir} < 0.1$); the remaining cells
are located in the CGM.  There is no {\OVI} absorption from the ISM,
only from the CGM.  We note that the phase properties of many of the
absorbing cells in the ISM do not satisfy our optically thin criteria
and should be viewed with caution, whereas all cells in the CGM do
meet the criteria.

For this simulated dwarf galaxy, we see that the majority of the CGM,
as selected by absorption with greater than 0.1~{\AA}, is
characterized by temperatures in the range $4 \leq \log T \leq 5$; the
absorbing gas is primarily what we might call the ``cool/warm'' CGM.
Interestingly, the gas that gives rise to {\OVI} absorption also
resides in this temperature range.  It would seem that the CGM of this
dwarf galaxy is primarily photoionized gas (in
Section~\ref{sec:ion-eq}, we quantify this and also discuss the
appropriateness of the assumption of equilibrium ionization modeling).
A minority of the absorbing cells have $\log T > 5$, and it is
possible that these cells are dominated by collisional ionization
processes.

The densities of the {\MgII} absorbing gas reside in the range $-3.5
\leq \log n\subH \leq -1.5$ and the temperatures are confined to a
narrow range at $\log T \simeq 4$.  Note that a substantial fraction
of the {\CIV} absorbing cells also reside in this temperature range.
However, the majority of the {\CIV} and {\OVI} absorbing gas primarily
has lower densities in the range $-6 \leq \log n\subH \leq -4$.

The relative population of absorbing cells in phase space is such
that {\HI} and {\CIV} absorption is selecting out significantly more
cells than {\MgII} and {\OVI} absorption.  To the degree to which the
CGM of this simulated dwarf galaxy reflects that of real-world dwarf
galaxies, this would immediately suggest the observed covering
fraction for {\HI} and {\CIV} absorption in dwarfs should be
substantially higher than the covering fraction for {\MgII} and {\OVI}
absorption.  Indeed, the covering fraction of {\CIV} absorption for $L
< 0.2L^{\ast}$ galaxies in the COS-Dwarfs sample is greater than 40\%
for $D/R_{\rm vir} \leq 0.5$ for a detection threshold of $\sim
0.1$~{\AA} \citep{bordoloi14}.  We defer study of how the covering
fractions and the equivalent width and column density impact parameter
distributions respond to different feedback recipes for future work
(Vander Vliet {\etal} 2014a, in preparation). 

The absorbing gas phase distributions for this simulated dwarf galaxy
show some similarities and some differences with those obtained by
\citet{ford13a, ford13b}, who present {\HI}, {\MgII}, and {\CIV}
absorption phase diagrams for a $z=0.25$ simulated galaxy with a halo
mass of $10^{12}$~M$_{\odot}$.  Their simulations are performed with
SPH, whereas we have used AMR simulations.

For {\HI} and {\MgII}, our absorbing gas phase distributions, obtained
for 1000 LOS with $D < 90$~kpc, are qualitatively consistent with the
phase distributions for the LOS with $D=100$ kpc and $D=1$ Mpc from
\citet{ford13a}.  This is likely because the overdensity of the CGM
within $D=100$ kpc in a dwarf galaxy halo is similar to that of a
$10^{12}$~M$_{\odot}$ halo at $100 < D \leq 1000$~kpc.  For {\CIV},
the temperature distribution for our dwarf is similar to that found by
\citet{ford13a}, however, the distribution of $n\subH$ in the CGM of
our dwarf galaxy peaks at $n\subH \sim 10^{-5}$ cm$^{-2}$, whereas,
the peak found by \citet{ford13a} is $n\subH \sim 10^{-4}$ cm$^{-2}$
for $100 < D \leq 1000$~kpc, and is $n\subH \sim 10^{-3}$ cm$^{-2}$
for $D = 10$~kpc. We cannot compare {\OVI} because \citet{ford13a} did
not present a $n\subH$--$T$ phase diagram for this ion.

\subsection{Kinematic and Spatial Relationships}
\label{sec:kin-space}

Here, we present insights into the application of observational
analysis techniques by examining the kinematic and spatial
relationship between the absorption profiles and the CGM gas giving
rise to absorption in simulations.  For our discussion, we again focus
on the {\HI}, {\MgII}, {\CIV}, and {\OVI} absorption for LOS 0092.

Of foremost interest (also see Figure~\ref{fig:los}) is that, whereas
the {\HI} absorption-selected cells are coincident with the {\MgII}
absorbing cells, these cells represent 15\% of the {\CIV} selected
cells and none of the cells selected by {\OVI} absorption.  This
non-coincidence of {\HI} and {\OVI} absorption-selected gas would not
be perceptible in real-world observations.  Since some of the {\HI}
absorption has LOS velocities coincident with the {\OVI} absorption,
it is likely that when employing common observational analysis methods
some fraction of the {\HI} column density would be attributed to the
{\OVI} absorbing gas phase (what fraction depends upon a somewhat
subjective educated estimate or ``artfully'' modeling of the
absorption, such as VP decomposition).  Therefore, it is almost
assured that the metallicity of the {\OVI} absorbing phase would then
be underestimated in observational studies.  

This would suggest that there may be systematic bias in the
metallicity determinations derived from observed {\OVI} systems that
exhibit low-ionization absorption conditions, since the {\HI} column
density might arise only within the low-ionization phase.  Note that,
to a large degree, the same argument holds for the {\CIV} absorption,
for which the majority of the absorbing cells are not selected by
{\HI} absorption.  In their SPH simulations, \citet{oppenheimer09}
also find that {\HI} and {\OVI} absorption does not trace the same
baryons, even though they are both present in the CGM.  They also
conclude that, if true, ths could result in underestimatations of the
true metallicity, since $\log Z/Z_{\odot} \propto (N_{\hbox{\tiny
    OVI}}/N_{\hbox{\tiny HI}})(f_{\hbox{\tiny HI}}/f_{\hbox{\tiny
    OVI}})$.

It is possible that the lack of coincident {\HI} and {\OVI} absorbing
gas could be a resolution effect in AMR simulations (and SPH
simulations).  In low density regions, where the grid-cell sizes are
larger, thermal and dynamic instabilities leading to condensations of
cooler higher density gas cannot be resolved.  Thus, multiphase
structures that might form embedded in the hotter lower density
regions of the simulated CGM would be suppressed.  Such structures may
have been detected observationally \citep{crighton14}.  
{\HI} and {\MgII} absorption arise exclusively in
cells with lengths of $\sim 3$ kpc, where as {\CIV} and {\OVI} arise
in cells with lengths ranging from 3--11 kpc and 6--45 kpc,
respectively.  Even in these extended cells, the ionization fraction
of {\HI} is so small that the path length cannot compensate to produce
detectable {\HI} absorption.  We note, that even if condensations were
resolved and successfully modeled in the simulations, they would still
comprise a separate phase from the {\OVI} absorbing gas and the {\HI}
and {\OVI} absorbing cells would not be coincident.  Thus, we caution
that {\OVI} absorption may rarely be associated with the same gas
structures giving rise to {\HI} absorption.

\begin{figure*}
\epsscale{0.90}
\plotone{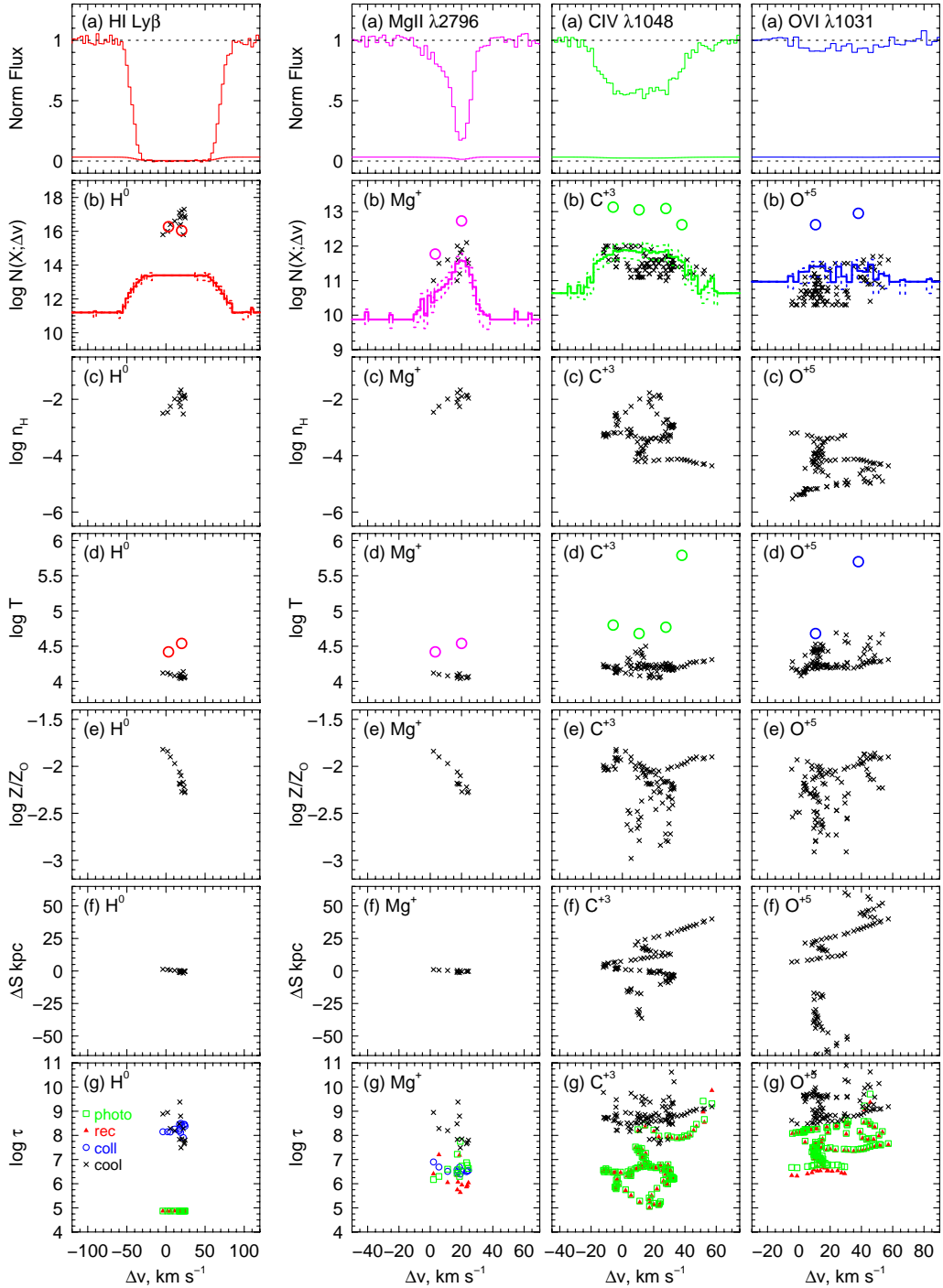}
\caption{(from top to bottom) Absorption profiles, column densities,
  hydrogen densities, temperatures, spatial locations, and ionization
  timescales of ``absorbing'' cells as a function of LOS velocity for
  LOS 0092.  (a) The absorption profiles as described in
  Figure~\ref{fig:profiles}.  (b) The AOD column densities, $N_a({\rm
    X}; \Delta v)$, (histograms) obtained from the spectra, the
  absorbing cell column densities (black points), and the VP component
  column densities (open circles).  (c) The absorbing cell hydrogen
  number densities, $n\subH$.  (d) The absorbing cell temperatures,
  $T$ (black points) and inferred VP component temperatures (open
  circles; the $\times$ corresponds to the column density upper
  limit).  (e) The absorbing cell metallicities, $\log Z/Z_{\odot}$.
  (f) The absorbing cell LOS positions, $\Delta S$.  (g) The absorbing
  cell photoionization (green), recombination (red), collisional
  (blue), and cooling timescales (black).}
\label{fig:AOD-times}
\end{figure*}

In Figure~\ref{fig:AOD-times}(a) we re-plot the absorption profiles.
In Figures~\ref{fig:AOD-times}(c), \ref{fig:AOD-times}(d), and
\ref{fig:AOD-times}(e) and we plot the absorbing cell hydrogen number
densities, $n\subH$, temperatures, $T$, and metallicities
$Z/Z_{\odot}$ in solar units as a function of LOS velocity.  What is
immediately notable is that the gas phases sampled by the
low-ionization ions cover a narrow range clustered around $\log n\subH
\simeq -2$ and $\log T\simeq 4.1$.  On the other hand, there is a
fairly steep metallicity gradient with LOS velocity, $\Delta \log
(Z/Z_{\odot})/\Delta v = -0.5/30 = -0.2$ dex ({\kms})$^{-1}$.  Note
that the velocity range over which the {\MgII} absorption is strongest
(where the highest column density cells are selected) coincides with
the lowest metallicity {\MgII} absorption selected cells.  The
relatively strong absorption is due to the slightly higher $n\subH$
and lower $T$ values, suggesting that the ionization balance dominates
the profile shape, not the metallicity.  Stronger absorption {\it
  within\/} a profile does not necessarily suggest higher metallicity.


In contrast, the gas phases of the higher-ionization ions represent a
much broader range of phases, $-4.5 \leq \log n\subH \leq -1.5$ and
$4.1 \leq \log T\leq 4.5$ for {\CIV} absorption selected gas and
$-5.5 \leq \log n\subH \leq -3.0$ and $4.1 \leq \log T\leq 4.7$ for
{\OVI} absorption selected gas.  Likewise, the metallicity ranges for
these ions is quite large, $-3.0 \leq \log Z/Z_{\odot} \leq -1.8$.
Unlike the low-ionization ions, there is no trend in these highly
variable quantities with LOS velocity.

In Figure~\ref{fig:AOD-times}(f), we plot the LOS spatial locations,
$\Delta S$, of the absorption-selected cells as a function of LOS
velocity.  Both the {\HI} and {\MgII} absorption arise in what might
be considered a ``cloud''.  By this term, we mean that the absorption
arises in spatially contiguous cells over a few kiloparsec LOS
pathlength and that there is little variation in the number densities
and temperatures of this gas over this short pathlength [see
  Figures~\ref{fig:AOD-times}(c) and \ref{fig:AOD-times}(d)].

The point here is that modeling the {\HI} and {\MgII} absorption with
VP decomposition, though not precisely appropriate, would not be
entirely without justice.  As shown in Figure~\ref{fig:vpfit}, the VP
profile model comprises two components, which reflect the velocity
structure of the {\MgII}, {\CII}, and {\SiII} absorption and seems
appropriate given the two clusters of low-ionization absorbing cells
in LOS velocity.  Furthermore, the very narrow range of cell
temperatures is consistent with the isothermal assumption inherent in
VP decomposition.

When we examine the physical properties of the cells selected by
{\CIV} and {\OVI} absorption, we see an entirely different physical
situation.  The densities and temperatures in these cells cover a
2.5 dex and 0.5 dex range, respectively.  The spatial locations are
spread out on the scale of 100 kpc.  Clearly, the gas selected
by higher-ionization absorption profiles for this simulation do not
arise in a cloud-like structure.  We revisit this below.

We now compare the AOD column density and VP fitting parameters to the
column densities of the absorption-selected cells.  The $N_a({\rm
  X}\supj;\Delta v)$ profiles are shown in
Figure~\ref{fig:AOD-times}(b). The dotted histograms are the
asymmetric $\pm 1~\sigma$ uncertainties in each velocity pixel.  The
horizontal lines outside the velocity range of the detected absorption
provides the upper limit on the AOD column density in the continuum of
the spectra.  Due to the black-bottomed saturation of the {\HI} {\Lyb}
absorption, the $N_a({\HI};\Delta v)$ profile provides a lower limit
on the column density in the velocity range $\Delta v \simeq \pm
40$~{\kms}.  The individual points (black $\times$ symbols) provide
the column densities of the absorption-selected cells for each
transition as a function of LOS velocity.  

Also shown are the VP component column densities (colored open
circles; see Figure~\ref{fig:vpfit} for the VP model profiles and
Table~\ref{tab:vpfit} for the VP component fitting parameters).  We
remind the reader that the VP models were segregated by ionization
level, with {\HI} and {\MgII} (along with {\CII} and {\SiII}) being
simultaneously fit for the lower ionization phase, and {\CIV} and
{\OVI} (along with {\CIII}) being simultaneously fit for the higher
ionization phase.  The low-ionization model is a two-component fit and
the high-ionization model is a four component fit.

Consider the {\MgII} absorption.  The column densities of a majority
of the cells {\it exceed\/} the $N_a({\MgII};\Delta v)$ profiles.
This is a fairly well understood expectation that is due to the
intrinsic absorption profiles from the gas being unresolved by the
spectrograph \citep[e.g.,][]{savage91} for which correction methods
have been developed \citep{jenkins96}.  For $\log T = 4.1$, the
intrinsic absorption profile width is $\simeq 6.6$ {\kms} (FWHM) for
{\MgII}, whereas the FWHM of the instrumental function for the
simulated spectra is $\simeq 17$~{\kms}.  Note that discrepancy
between $N_a({\rm X}\supj;\Delta v)$ and the cell column densities is
also seen for the coolest gas selected by {\CIV} and {\OVI}
absorption, since these cells give rise to the narrowest intrinsic
line shapes.

The VP component column densities for the {\HI} absorption are fairly
consistent with the cell column densities, though the $\Delta v =
20$~{\kms} component is $\simeq 1$ dex below the dominant absorbing
cells.  The column densities of the {\MgII} VP components are also
$\simeq 0.5$ dex higher than the peak column densities of the
absorbing cells.  If we examine the inferred temperatures of the gas
from the VP components [Figure~\ref{fig:AOD-times}(d)], we see that
the component temperatures are also overestimated.

Since the lower ionization profiles are decomposed into only two
components, the Doppler $b$ parameters are likely to be systematically
large; that is, the kinematic morphology of the absorption is due to a
steep velocity gradient across the LOS (see the $\Delta v$ panels of
Figure~\ref{fig:los}), which has been modeled as two thermally
broadened ``clouds''.

For the {\CIV} and {\OVI} absorption, the VP component column
densities and temperatures are significantly overestimated, roughly by
$1.5$ dex and $\simeq 0.5$ dex, respectively.  In fact, the inferred
temperatures from the VP components are $\log T \simeq 4.7$, and are
approaching temperatures commonly assumed to be indicative of
collisional ionization for these two ions (in the following
subsection, we show that the {\CIV} and {\OVI} absorbing gas is
dominated by photoionization equilibrium).  The very broad component
suggests $\log T \simeq 5.7$, but the VP component broadening is
clearly modeling the bulk velocity flow.  These considerations serve
as a warning that VP decomposition can be highly misleading about the
ionization condition of the gas, and may simply be a flawed approach
to modeling {\CIV} and {\OVI} absorbing gas (as we discuss further
below).

These discrepancies with the VP decomposition would undoubtedly
systematically skew estimated of the gas density and metallicity
inferred from ionization modeling of the VP components.  However, when
we examine the {\it total\/} column densities, we find good agreement
between the AOD columns, $N_a(X\supj)$, the VP column densities, and
the absorbing cell column densities.

In Table~\ref{tab:cols}, we list the total column densities for {\HI},
{\MgII}, {\CIV}, and {\OVI}.  The VP total column density is the sum
of the components, the AOD total column density is given by
Eq.~\ref{eq:Naod}, and the total for the ``Cells'' is the sum of the
absorption selected cell column densities.  That the totals are in
good agreement is promising, for the consistency between observational
methods for obtaining the total column densities and the actual total
column density of the gas selected by absorption would imply that mass
estimates of the CGM using integrated AOD profiles and summed VP
column densities accurately recover the gas mass for both the low and
high ionization CGM \citep[for example, as done by][]{tumlinson11,
  bordoloi14}.

\begin{deluxetable}{lrrr}
\tablecolumns{4}
\tablewidth{0pt}
\setlength{\tabcolsep}{0.06in}
\tablecaption{Comparison of Total Column Densities\tablenotemark{a}
\label{tab:cols}}
\tablehead{
  \colhead{(1)}  &
  \colhead{(2)}  &
  \colhead{(3)}  &
  \colhead{(4)}   \\[2pt]
  \colhead{Ion}  &
  \colhead{$N_{\rm tot}({\rm VP})$}  &
  \colhead{$N_{\rm tot}({\rm AOD})$}  &
  \colhead{$N_{\rm tot}({\rm Cells})$}  \\
  \colhead{}      &
  \colhead{(cm$^{-2}$)}        &
  \colhead{(cm$^{-2}$)}        &
  \colhead{(cm$^{-2}$)}        
}
\startdata
{\HI}   &  $16.47\pm0.10$ &             $>14.69$ & $17.95$ \\[3pt] 
{\MgII} &  $12.78\pm0.01$ & $12.75^{+0.08}_{-0.06}$ & $12.76$ \\[5pt] 
{\CIV}  &  $13.61\pm0.06$ & $13.62^{+0.05}_{-0.03}$ & $13.57$ \\[5pt] 
{\OVI}  &  $13.13\pm0.42$ & $13.02^{+0.08}_{-0.03}$ & $13.03$  
\enddata
\tablenotetext{a}{The total column densities of the absorbing gas for
  each analysis method, where $N_{\rm tot}({\rm VP})$ is the sum of
    the individual VP component column densities, $N_{\rm tot}({\rm
      AOD})$ is the integrated AOD column density given by
    Eq.~\ref{eq:Naod}, and $N_{\rm tot}({\rm Cells})$ is the sum of
    the cell column densities along the LOS that contribute to the
    absorption profiles.}
\end{deluxetable}

Interpretation of both AOD column density profiles and VP
decomposition is predicated on the assumption that the absorption at a
given LOS velocity arises from the same physical gas complex, or
cloud.  Whereas the $N_a({\MgII};\Delta v)$ profile reasonably
recovers the distribution of {\MgII} absorbing gas column densities
with LOS velocity, the same cannot be said for the gas giving rise to
{\CIV} and {\OVI} absorption.  For these ions, we see that multiple
absorbing gas structures arise with virtually identical velocities and
that these structures can have up to a $\sim 1$ dex spread in column
densities.  That is, the flux decrement at any given velocity can
result from the sum of the column densities from multiple cells that
happen to have the same LOS velocity.

When we further consider the spatial locations of absorbing cells with
the same LOS velocity, we find that the assumption underlying standard
observational analysis of the absorption profiles is not validated by
the kinematic-spatial distribution of the absorbing gas for all
ions.  As mentioned above while discussing
Figure~\ref{fig:AOD-times}(f), the {\HI} absorption and the {\MgII}
absorption is distributed over a range of velocities in gas that
forms a single contiguous structure at $\Delta S \simeq 0$ (i.e., a
single ``cloud'').  However, at most LOS velocities, the {\CIV} and
{\OVI} absorbing gas originates from multiple groupings of cells
with physical separations ranging from a few to 150 kpc.  

For example, the {\CIV} absorption in the range of $10 \leq \Delta v
\leq 20$~{\kms} arises in six physically distinct LOS locations spread
over 100 kpc with typical LOS separations of $\sim 20$ kpc, whereas the
cells in each of these spatial groupings are contiguous over one
to a few kpc.  Not only does the {\CIV} absorption arise in gas
distributed over $\sim 100$ kpc, the gas is characterized by a complex
velocity field that reverses direction along the LOS several times
that results in LOS velocity caustics where the column densities of
physically unconnected gas structures are summed.  The {\OVI}
absorption exhibits similar behavior.  We note that in a double
line-of-sight experiment, \citet{muzahid14} concluded that the {\OVI}
absorption arises in large extended structures that are consistent with
our simulation results.  The fact that {\OVI} and {\CIV} arise in
these extended structures with complex spatial and velocity fields is
no doubt the reason that the VP component Doppler $b$ parameters yield
inferred temperatures that are $\sim 1$ dex too high [see
  Figure~\ref{fig:AOD-times}(d)].

In summary, current observational analysis techniques may be valid for
low-ionization gas.  As mentioned above, the {\HI} and {\MgII}
absorption arises in a single contiguous gas structure.  The density
and temperature ranges of this structure are $\log n\subH \simeq -2$
and $\log T \simeq 4.1$, respectively.  Based upon standard
observational techniques and Cloudy modeling applied to COS spectra,
\citet[][$L \sim L^{\ast}$ galaxies]{werk14} and
\citet[][$L>0.1L^{\ast}$ galaxies]{stocke13} constrained the
properties of the cool/warm photoionized CGM to reside near these
values and to be generally consistent with the phase diagrams of
{\MgII} and {\HI} absorbing gas shown in Figure~\ref{fig:phases}.
However, becasue of the complex spatial-kinematic distrbution of the
higher-ionization gas, current observational techniques of {\CIV} and
{\OVI} absorption likely do {\it not\/} correctly model the underlying
physics of the absorbing gas.

The implication of the kinematic-spatial distribution of the absorbing
gas is that observational analysis techniques (i.e., AOD column
density and/or VP decomposition coupled with ionization modeling) that
are founded on the absorbing gas at a given LOS velocity arising in a
physically contiguous gas structure may be quite invalid (especially
for high-ionization species).  It remains to be determined to what
degree of accuracy the inferred densities, temperatures, and
metallicites from observational modeling techniques are recovered when
applied to the simulated spectra and compared to the gas cell
properties.  In the near future, we aim to undertake such a study
(Vander Vliet {\etal} 2014b, in preparation). 

\subsection{Ionization Equilibrium}
\label{sec:ion-eq}

We now turn to the question as to whether CGM absorbing gas can be
modeled with equilibrium ionization models.  We also investigate
methods for determining if the absorbing gas is photoionized or
collisionally ionized.  

Since most all ionization modeling of observational data is based upon
the assumption of ionization equilibrium, it is important to
understand to what degree this assumption holds.  Our ionization
modeling, and therefore all subsequent absorption line analysis of the
simulations, is also is based upon this assumption.  Investigations
into non-equilibrium ionization had been investigated by \citet{cen06}
and by \citet{oppenheimer13}.  Non-equilibrium collisional ionization
results in a reduction of the {\OVI} column density relative to the
equilibrium solution, though \citet{oppenheimer13} find that
non-equilibrium effects are not as pronounced when photoionization is
taken into account.

Whether the gas is predominantly photoionized or collisionally ionized
has implications for correctly estimating the density, temperature,
physical size, cooling time, and estimated mass of the absorbing gas
(in otherwords, understanding the physical nature of the CGM, and
subsequently its role in galaxy evolution).  In observational work one
cannot directly deduce whether the gas is dominated by photoionization
or collisional ionization. Often, inferences of collisional
iopnization are based upon the presence of large thermal velocity
widths in the absorption profiles, or the column density ratios of
different ions in different ionization stages \citep[e.g.,][]{tripp01,
  simcoe06, tripp11, fox14}.  However, from the ionization modeling
one can directly examine the degree to which the ionization balance in
the absorbing gas is dominated by photoionization or collisional
ionization.  The method we employ can also be applied to observational
work.

The condition of ionization equilibrium is that the gas must be
thermally stable over the time required for the photoionization,
recombination, and collisional ionization rates to achieve a
steady-state balance.  If the time required to achieve ionization
equilibrium is denoted $\tau_{\hbox{\tiny eq}}$, the condition is
often expressed,
\begin{equation}
\tau_{\hbox{\tiny eq}} \ll \tau\subcool \, ,
\label{eq:eqcool}
\end{equation}
where 
\begin{equation}
\tau\subcool = \frac{Q}{\left| dQ/dt \right|} 
= \frac{Q}{\left| \Lambda_{\hbox{\tiny net}} \right|} 
= \frac{\frac{3}{2}\left( \eden + n\subN \right) kT} {\left|
  \Lambda_{\hbox{\tiny heat}} - \Lambda_{\hbox{\tiny cool}} \right|}
\quad {\rm sec} \, ,
\label{eq:coolingtime}
\end{equation}
is the cooling timescale.  Here, $Q$ is the energy density of the gas
[erg cm$^{-3}$], $\eden$ is the electron density, $n\subN$ is the
total number density of all ions, $\Lambda_{\hbox{\tiny heat}}$ is the
heating rate per unit volume, and $\Lambda_{\hbox{\tiny cool}}$ is the
cooling rate per unit volume [erg s$^{-1}$ cm$^{-3}$].  Both
$\Lambda_{\hbox{\tiny heat}}$ and $\Lambda_{\hbox{\tiny cool}}$ are
functions of the spectral energy distribution of ionizing photons, the
gas density, $n\subH$ temperature, $T$, and the relative abundances of
the metals.

The ionization timescales are different for each ion in the gas.  As
such, it is possible that some ions are in ionization equilibrium,
while other ions are not.  For our work, which focuses on the
absorption properties of ions that are commonly observed in absorption
lines studies, we aim to determine if the assumption of ionization
equilibrium applies for these ions in particular.

Here, we examine three timescales, photoionization, $\tau\ph $,
recombination, $\tau\rec$, and collisional ionization, $\tau\coll$,
for the four ions H$^{\hbox{\tiny 0}}$, Mg$^{+}$, C$^{\hbox{\tiny
    +3}}$ and O$^{\hbox{\tiny +5}}$ in the absorption selected gas
shown in Figures~\ref{fig:los} and \ref{fig:AOD-times} for LOS 0092.

The ionization timescales for ion X$\supj$ are obtained by rearranging
the rate equations \citep[see][for full details]{cwc-ioncode} into the
form $(1/n\subXj)(dn\subXj/dt) = R\subXj$, where $R\subXj$ is the rate
[s$^{-1}$] at which $n\subXj$ is either increasing or decreasing.  The
solution is of the form $e^{-t/\tau\subXj}$.  Thus, the ionization
timescales are the $e$-folding time for a change in the numder
density.  The rate $R\subXj$ is a function of the ionizing photon
field, $J_{\nu}$, ion densities, electron density, and gas
temperature, which are all assumed to change by no more than a
negligible amount so that $R\subXj$ can be approximated as a constant
(thus, greatly simplifying the integrals; as such, the timescales are
approximations).  Integrating over the time interval $\tau\subXj = t_1
- t_0$ with the condition $n\subXj(t_1)/n\subXj(t_0) = e^{\pm1}$
yields $\tau\subXj = \left| R\subXj \, \right| ^{-1}$.  The absolute
value reflects the fact the change in $n\subXj$ may be decreasing or
increasing over the time interval, but we wish to know the timescale
only, regardless of the sign of $dn\subXj/dt$.

Accounting only for photoionization rates (omitting Auger ionization),
we obtain the photionization timescale,
\begin{equation}
\tXj\ph \simeq \left| \, R\ph\subXj 
- \frac{n\subXjm}{n\subXj} R\ph\subXjm \right| ^{-1} \quad {\rm sec} \, ,
\label{eq:tau-photo}
\end{equation}
where $R\ph\subXj(J_\nu)$ is the photoionization rate [s$^{-1}$] for
ion X$\supj$, and the density units are cm$^{-3}$.  In the case of the
neutral ion ($j=1$), the rates indexed by $j-1$ do not enter into the
derivation and are omitted from Eq.~\ref{eq:tau-photo}.

Accounting for both photo and dielectronic recombination, we obtain
the recombination timescale,
\begin{equation}
\tXj\rec \simeq \left| \, \eden \left[ 
\left( \beta\phr\subXjm  + \beta\die\subXjm \right)
  - \frac{n\subXjp}{n\subXj} 
\left( \beta\phr\subXj  + \beta\die\subXj \right)
\right] \right| ^{-1} \quad {\rm sec} \, ,
\label{eq:tau-recomb}
\end{equation}
where $\beta\phr\subXjm(T)$ and $\beta\die\subXjm(T)$ are the photo
and dielectronic recombination rate coefficients [cm$^3$~s$^{-1}$] for
ion X$\supj$.

Accounting for both direct and excitation-auto collisional ionization
processes, we obtain the collisional ionization timescale\footnote{In
  some applications, the collisional timescale includes the balance
  with reombination.  Here, we desire to know the collisional
  ionization timesale only.  If the timescale for the balancing of
  recombination and ionization is desired, it can be computed from
  $\tXj\rec \tXj\coll/(\tXj\rec + \tXj\coll)$ using
  Eqs.~\ref{eq:tau-recomb} and \ref{eq:tau-coll}.},
\begin{equation}
\tXj\coll \simeq \left| \, \eden 
\left[ 
\left( \alpha\cdi\subXj +  \alpha\cea\subXj \right) - 
\frac{n\subXjm}{n\subXj} 
\left( \alpha\cdi\subXjm +  \alpha\cea\subXjm \right)  
\right] \right| ^{-1} \quad {\rm sec} \, ,
\label{eq:tau-coll}
\end{equation}
where $\alpha\cdi\subXj(T)$ and $\alpha\cea\subXj(T)$ are the direct
and excitation-auto ionization rate coefficients [cm$^3$~s$^{-1}$] for
ion X$\supj$.  For the derivation of Eq.~\ref{eq:tau-coll}, we have
omitted charge exchange reactions.

For each cell in the simulation, we compute the energy density, $Q$,
and the photoionization, recombination, and collisional ionization
timescales directly from the photoionization rates, equilibrium
electron density, and the recombination and collisional ionization
rate coefficients.  The computation of the rates, rate coefficients,
and the equilibrium ionization condition for each gas cell is
described in \citet{cwc-ioncode}.

To compute the cooling timescale, we adopt the cooling functions of
\citet{wss09}.  These cooling functions were generated with Cloudy
07.02 \citep{ferland98} for model clouds as a function of hydrogen
number density, temperature, metallicity, and redshift.  The ionizing
radiation is the \citet{haardt01} ultraviolet background (UVB).  The
clouds are assumed to be dust free and optically thin.  As with our
ionization model, the gas is assumed to be in ionization equilibrium.
We note here that it is important the cooling function accounts for
photoionization processes.  At a given gas density and temperature,
photoionization will yield a gas that is more highly ionized than
collisionally ionized gas.  This results in cooling rates that are
diminished as compared to the rates for collisional ionized gas
\citep[][and references therein]{wss09, schure09}. Thus, the standard
collisional ionization equilibrium cooling functions
\citep[e.g.,][]{sutherland93} can substantially understimate the
cooling timescales in photoionized gas.

We also note that the \citet{wss09} cooling functions are highly
consistent with the cooling functions used by hydroART
\citep{ceverino09, st13-dwarfs, Ceverino14}. This is important because
the cooling functions dictate the equilibirum temperature of the gas,
and the gas temperature determines the collision based ionization and
recombination rates in equilibrium ionization models.

The cooling functions are determined by interpolating across the
high-resolution tables of \citet{wss09}, which return the normalized
net cooling function, $\Lambda_{\hbox{\tiny net}}/n^2\subH$.  We
determined $\left| \Lambda_{\hbox{\tiny net}} \right|$ for each gas
cell in the simulation box.  The cell data required for the
interpolation are the hydrogen number density, $n\subH$, temperature,
$T$, redshift, $z_{\rm box}$ (needed for the UVB), and the abundances of
He, C, N, O, Ne, Mg, Si, S, Ca, and Fe, relative to hydrogen.  The
latter are detemined from the type II and Ia supernovae mass fractions
in the gas cell.  Note that the {\it individual\/} contributions of
the aforementioned ions, based on their number densities relative to
hydrogen, are accounted for in the cooling functions.  We then apply
Eq.~\ref{eq:coolingtime} to compute $\tau\subcool$ for each cell in
the simulation box.

In Figure~\ref{fig:AOD-times}(g), we plot the timescales for
photoionization (green open squares), recombination (red filled
triangles), and collisional ionization (open blue circles) for the
four ions H$^{\hbox{\tiny 0}}$, Mg$^{+}$, C$^{\hbox{\tiny +3}}$, and
O$^{\hbox{\tiny +5}}$ in the absorption selected cells. We also
plot the cooling timescales (black $\times$ symbols).

The H$^{\hbox{\tiny 0}}$ ion is clearly in photoionization equlibrium.
The photoionization and recombination timescales are equal and both of
these are roughly three orders of magnitude shorter than the
collisional ionization timescales.  Furthermore, the cooling
timescales are also roughly three orders of magnitude longer than the
photoionization timescales, which satisfies the criterion given by
Eq.~\ref{eq:eqcool} for ionization equilibrium.  The collisional
ionization processes of hydrogen do not achieve equilibrium, since the
gas cools on similar timescales, but photoionization completely
dominates the ionization of hydrogen.  The Mg$^{+}$ ion, on the
otherhand, is in near balance between photoionization, collisional
ionization, and recombination.  However, the assumption of ionization
equilibrium is well satisfied for the majority of gas in that
the cooling timescales exceed the ionization timescales.

In the cases of C$^{\hbox{\tiny +3}}$ and O$^{\hbox{\tiny +5}}$, the
ions are in equilibrium between the recombination and photoionization
timescales.  For both ions, the collisional ionization timescales
exceed $100$ Gyr; the gas giving rise to {\CIV} and {\OVI}
absorption is clearly dominated by photoionization.  Furthermore, the
assumption of ionization equilibrium is well supported by the long
cooling timescales.  At LOS velocities of $\Delta v \simeq 50$~{\kms},
the gas may be only marginally in ionization equilibrium for these
ions.


In summary, the assumption of ionization equlibrium for the commonly
observed ions H$^{\hbox{\tiny 0}}$, Mg$^{+}$, C$^{\hbox{\tiny +3}}$,
and O$^{\hbox{\tiny +5}}$ is sound for modeling the ionization
conditions of the absorbing gas (at least for the CGM of this
simulated galaxy).  This lends a great deal of confidence that
observational data can be accurately analyzed using ionization
equilibrium for these ions.  Futhermore, we have shown that it is
straight forward to determine whether a given ion is photoionized or
collisionally ionized in the absorbing gas of the simulated CGM.  In
principle, the method can be applied to observational data if the
photoionization, collisional ionization, and recombination rate
coefficients from the ionization modeling are retreivable.  However,
we caution that using column densities of individual VP components may
not yield an ionization model that reflects the absorbing gas
conditions (see Section~\ref{sec:kin-space}).

\section{Conclusions}
\label{sec:conclude}

We have presented a case as to why a comprehensive understanding of
the CGM via absorption line studies requires objective absorption line
analysis of high resolution hydrodynamic cosmological simulations that
mirrors observational absorption line methods.  Most importantly, we
argued and demonstrated that the synthetic absorption line spectra of
the simulated CGM should emulate the observed spectra in all of its
characteristics, as should the analysis of the absorption line
measurements obtain directly from the spectra.  Primarily, this
assures that the gas selected by absorption in the simulated CGM is
directly comparable to the gas selected by absorption in the observed
CGM.  The point is that, when comparing the simulated CGM gas
properties (densities, temperatures, kinematics, ionization and
chemical conditions, and spatial distributions) to the properties of
the CGM gas inferred from observational absorption line studies, only
the gas that is selected by the simulated absorption lines should be
compared.  Only in this fashion are direct insights into observational
analysis and the physical nature of the CGM gleaned.  

We generated absorption profiles for {\HI} the {\Lya} and {\Lyb}
transitions, and for the spectra of {\HI} {\Lya} and {\Lyb}, {\SiII}
$\lambda \lambda 1190, 1193$, {\SiII} $\lambda 1260$, {\CII} $\lambda
1036$ and $\lambda 1335$, {\MgIIdblt}, {\CIVdblt}, and {\OVIdblt}
metal-line transitions for 1000 LOS through the CGM of a simulated
dwarf galaxy from the work of \citet{st13-dwarfs}.  Though in this
paper we focused on a single LOS for illustrating the relationships
between commonly observed absorption lines and the underlying
properties of the absorbing gas, the results discussed for this LOS
are generally true from LOS to LOS.  Highlights of our findings for
the CGM of this simulated galaxy are:

\vglue 0.1in
(1) The simulations indicate that low ionization gas, as probed in
    absorption with ions having ionization potentials near the
    hydrogen ionization edge, likely arise from what might be called
    ``clouds''.  Along the LOS, these structures are characterized by
    a narrow range of densities and temperatures, suggesting they can
    be modeled as single phase structures.  In addition, the
    absorption lines arise in gas that is localized along the LOS
    (the grid cells in the simulations are spatially contiguous).  As
    such, commonly applied observational analysis methods that
    incorporate AOD column densities, VP decomposition, and
    single phase ionization modeling are likely to be sound.

\vglue 0.1in
(2) The simulations indicate that higher ionization gas likely arises
    in extended structures that sample gas distributed up to 50--100
    kpc along the LOS.  Furthermore, the absorption at a given LOS
    velocity can arise in several smaller scale regions of the gas
    separated by tens of kiloparsecs that circumstantially align in
    LOS velocity.  The gas phases (densities and temperatures) that
    give rise to absorption at a given LOS velocity can vary up to an
    order of magnitude or more. As such, AOD profiles and VP
    decomposition do not properly reflect the high-ionization
    absorbing gas properties.  Ionization modeling of higher
    ionization gas as ``cloud''-like structures should be viewed with
    healthy skepticism. However, AOD profiles and VP decomposition can
    accurately reflect the total gas mass and average gas properties,
    because the {\it total\/} column densities from these methods are
    highly consistent with the total gas column along a LOS.

\vglue 0.1in
(3) Higher ionization gas, such as {\OVI} absorption selected gas, can
    have LOS absorption velocities that overlap with the LOS
    absorption velocity of {\HI} absorption selected gas, but the
    simulations indicate that in not all cases is the {\HI} absorption
    associated with the higher ionization absorbing gas.  In addition
    to the caveats for point (2) above, this presents further
    challenges for ionization modeling and metallicity estimates of
    the high ionization CGM.

\vglue 0.1in 
(4) Estimates of the ionization timescales and the cooling timescales
of the gas that gives rise to detected absorption indicate that the
gas can be successfully modeled as being in ionization equilibrium.
Broad absorption in high-ionization ions does not necessarily indicate
high temperature gas that is often taken to imply collisional
ionization dominates.  In fact, we found that VP decomposition of
{\CIV} and {\OVI} absorption yielded temperatures consistent with
collisional ionization when in fact the {\OVI} absorbing gas is
roughly 1 dex cooler and the broadening is due to the complexity of
the spatial and velocity fields over the extended absorbing structure.
Indeed, analysis of the ionization modeling showed that the {\CIV} and
{\OVI} absorbing gas is photoionized.

\vglue 0.1in
To the extent that the CGM of this simulated galaxy reflects the real
world, the points we have discussed here provide our first qualitative
insights into the effectiveness or possible misapplication of standard
absorption line analysis methods applied to studies of the CGM.

However, we have studied the simulated CGM of a single isolated
low-mass (dwarf) galaxy.  The star formation history is highly
stochastic and ``bursty'' \citep{st13-dwarfs}, and there are no
satellite galaxies and filamentary accretion is negligible.  The
effects of both these characteristics are certainly manifest in the
global CGM properties of this simulated galaxy.  It remains to be seen
if the CGM of higher mass galaxies in more complex cosmological
environments presents a substantially different or broader set of
insights.

In the future, we aim to undertake a comprehensive quantitative
analysis of the simulated CGM, comparing the inferences derived from
various observational works to the inferences derived from the
application of observational analysis methods to simulated absorption
data.  Examples of observational data that can be compared are the
covering fractions, the column density and equivalent width impact
parameter distributions, and the kinematics.  In particular, these
quantities may promise to provide insight into how the CGM properties
reflect different stellar feedback recipes.

Using the absorption selected gas properties, we aim to also quantify
the degree to which the observational analysis methods accurately
recover the ``true'' properties of absorbing gas.  We aim to
determined to what degree of accuracy the inferred densities,
temperatures, and metallicities from observational modeling techniques
are recovered when applied to the simulated spectra and compared to
the gas properties.  This work will require full application of
the AOD column density, VP decomposition, and ionization
modeling techniques.  It is also an immediate goal to include and test
shielding and basic radiative transfer effects in our ionization model
\citep{cwc-ioncode} so that we can relax the optically thin
constraint.

\acknowledgments

We thank the referee for helpful comments that improved this
manuscript.  CWC, ST-G, and AK were partially supported through grants
HST-AR-12646 and HST-GO-13398 provided by NASA via the Space Telescope
Science Institute, which is operated by the Association of
Universities for Research in Astronomy, Inc., under NASA contract NAS
5-26555.  GGK was supported by an Australian Research Council Future
Fellowship FT140100933.  JRV acknowledges support through a NASA New
Mexico Space Grant Consortium (NMSGC) Graduate Research Fellowships.

\end{document}